\documentclass[12pt]{amsart}
\usepackage{amsmath,amsthm,amsfonts,amssymb,eucal,ifpdf,bbm}
\ifpdf
\usepackage{hyperref}
\usepackage{pdfsync}
\else
\usepackage[hypertex]{hyperref}
\fi
\renewcommand {\a}{ \alpha }
\renewcommand{\b}{\beta}

\newcommand{\g}{\gamma}

\newcommand{\vark}{\varkappa}

\renewcommand{\l}{\lambda}
\renewcommand{\L}{\Lambda}
\newcommand{\z}{\zeta}

\newcommand{\om}{\omega}
\newcommand{\Om}{\Omega}

\newcommand{\oq}{\ {\raise 7pt\hbox{${\scriptstyle\circ}$}}
\kern -7pt{%\lower 2pt
\hbox{$Q$}}}

\newcommand{\R}{ \mathbb R}
\newcommand{\N}{ \mathbb N}
\newcommand{\C}{\mathbb C}

\newcommand{\Rd}{ \mathbb R^d}

\newcommand {\GF}{\mathfrak F}

\newcommand {\GV}{\mathfrak V}
\newcommand {\GW}{\mathfrak W}
\newcommand {\GU}{\mathfrak U}

\newcommand {\GX}{\mathfrak X}
\newcommand {\ba}{\mathbf a}

\newcommand {\BD}{\mathbf D}

\newcommand {\BM}{\mathbf M}

\newcommand {\BS}{\mathbf S}

\newcommand {\bx}{\mathbf x}

\newcommand {\be}{\mathbf e}

\newcommand {\by}{\mathbf y}

\newcommand {\bs}{\mathbf s}

\newcommand {\bv}{\mathbf v}
\newcommand {\bn}{\mathbf n}

\newcommand {\bnu}{\boldsymbol\nu}

\newcommand {\bmu}{\boldsymbol\mu}
\newcommand {\bth}{\boldsymbol\theta}
\newcommand {\Bth}{\boldsymbol\Theta}
\newcommand {\boldeta}{\boldsymbol\eta}

\newcommand {\bphi}{\boldsymbol\phi}
\newcommand {\bxi}{\boldsymbol\xi}

\newcommand {\BXi}{\boldsymbol\Xi}
\newcommand {\Bxi}{\boldsymbol\Xi}

\newcommand{\bchi}{\boldsymbol\chi}

\newcommand{\lu}{\langle}
\newcommand{\ru}{\rangle}

\newcommand{\CV}{\mathcal V}

\newcommand{\CB}{\mathcal B}

\newcommand{\CX}{\mathcal X}

\newcommand{\CP}{\mathcal P}

\newcommand{\CA}{\mathcal A}

\newcommand{\CM}{\mathcal M}

\newcommand{\CC}{\mathcal C}

\newcommand{\CG}{\mathcal G}
\newcommand{\CW}{\mathcal W}

\newcommand{\plainC}[1]{\textup{{\textsf{C}}}^{#1}}
\newcommand{\plainB}{\textup{{\textsf{B}}}}
\newcommand{\plainS}{\textup{{\textsf{S}}}}

\newcommand{\plainH}[1]{\textup{{\textsf{H}}}^{#1}}

\newcommand{\plainL}[1]{\textup{{\textsf{L}}}_{#1}}

\DeclareMathOperator{\tr}{{tr}}
\DeclareMathOperator{\card}{{card}}

\newcommand{\1}
{{\,\vrule depth3pt height9pt width1pt}\,}

\DeclareMathOperator {\dist} {{dist}}

\DeclareMathOperator \vol{{vol}}

\DeclareMathOperator{\op}{{Op}}

\DeclareMathOperator {\ad}{{ad}}
\DeclareMathOperator{\supp}{{supp}}

\DeclareMathOperator{\Id}{\mathbbm 1}
\newtheorem{thm}{Theorem}[section]
\newtheorem{cor}[thm]{Corollary}
\newtheorem{cla}[thm]{Claim}
\newtheorem{lem}[thm]{Lemma}
\newtheorem{prop}[thm]{Proposition}

\theoremstyle{definition}
\newtheorem{defn}[thm]{Definition}%[section]

\newtheorem{rem}[thm]{Remark}

\numberwithin{equation}{section}

\newcommand{\bee}{\begin{equation}}
\newcommand{\ene}{\end{equation}}
\newcommand{\bees}{\begin{equation*}}
\newcommand{\enes}{\end{equation*}}
\newcommand{\bes}{\begin{split}}
\newcommand{\ens}{\end{split}}

\newcommand{\bet}{\begin{thm}}
\newcommand{\ent}{\end{thm}}
\newcommand{\bel}{\begin{lem}}
\newcommand{\enl}{\end{lem}}
\newcommand{\bec}{\begin{cor}}
\newcommand{\enc}{\end{cor}}
\newcommand{\becl}{\begin{cla}}
\newcommand{\encl}{\end{cla}}
\newcommand{\bep}{\begin{proof}}
\newcommand{\enp}{\end{proof}}
\newcommand{\ber}{\begin{rem}}
\newcommand{\enr}{\end{rem}}
\newcommand{\ep}{\varepsilon}
\newcommand{\la}{\lambda}
\newcommand{\La}{\Lambda}

\newcommand{\al}{\alpha}
\newcommand{\Z}{\mathbb Z}

\newcommand {\BUps}{\boldsymbol\Upsilon}

\newcommand{\CF}{\mathcal F}
\renewcommand{\ge}{\geqslant}
\renewcommand{\le}{\leqslant}

\makeatletter
\def\square{\RIfM@\bgroup\else$\bgroup\aftergroup$\fi
  \vcenter{\hrule\hbox{\vrule\@height.6em\kern.6em\vrule}\hrule}\egroup}
\makeatother

\begin{document}

\hoffset -4pc

\title[Asymptotic expansion of IDS]
{Complete asymptotic expansion of the integrated density of states of multidimensional almost-periodic pseudo-differential operators}
\author[S. Morozov, L. Parnovski \& R. Shterenberg]
{Sergey Morozov, Leonid Parnovski \& Roman Shterenberg}
\address{Department of Mathematical Sciences\\ Aarhus University\\ Ny Munkegade 118\\ DK-8000 Aarhus C\\ Denmark}
\email{smorozov@imf.au.dk}
\address{Department of Mathematics\\ University College London\\ Gower Street\\ London\\ WC1E 6BT\\ UK}
\email{leonid@math.ucl.ac.uk}
\address{Department of Mathematics\\ University of Alabama at Birmingham\\ 1300 University Blvd.\\
Birmingham AL 35294\\ USA}
\email{shterenb@math.uab.edu}


\date{\today}

\begin{abstract}
We obtain a complete asymptotic expansion of the integrated density of states of operators of the form $H= (-\Delta)^w+ B$ in $\R^d$. Here $w> 0$, and $B$ belongs to a wide class of almost-periodic self-adjoint pseudo-differential operators of order less than $2w$. In particular, we obtain such an expansion for magnetic Schr\"odinger operators with either smooth periodic or generic almost-periodic coefficients.
\end{abstract}

\

\maketitle
\vskip 0.5cm

\renewcommand{\uparrow}{{\mathcal {L F}}}
\newcommand{\ssharp}{{\mathcal {L E}}}
\renewcommand{\natural}{{\mathcal {NR}}}
\renewcommand{\flat}{{\mathcal R}}
\renewcommand{\downarrow}{{\mathcal {S E}}}

\renewcommand{\theenumi}{\alph{enumi}}

\section{Introduction}

In \cite{ParSht2}, two of the authors of this paper have obtained
the complete power asymptotic expansion of the integrated density of
states of Schr\"odinger operators
\bees
H= -\Delta +V
\enes
acting in $\R^d$ assuming that the real-valued potential $V$ is either smooth
periodic, or generic quasi-periodic, or belongs to a reasonably
wide class of almost-periodic functions (see \cite{ParSht2} for
a complete set of conditions on $V$ as well as the previous history
of the subject). The main aim of the current paper is to extend the results of \cite{ParSht2} to a more general class of operators.
We give a detailed description of this new class in the next section; here, we list the main properties of the operators belonging to it.

(i) We consider perturbations of the Laplacian, or any positive power of the Laplacian. More precisely, we work with operators of the form
\bees
H= (-\Delta)^w +B,
\enes
where $B$ is a differential or pseudo-differential operator of order $\kappa<2w$. Here $H$ is self-adjoint and belongs to the standard algebra of almost-periodic pseudo-differential operators, see e.g. \cite{Shu0} and \cite{Shu}.

(ii) If $B$ is a differential operator, we assume that its coefficients satisfy the same conditions the potential $V$ had to satisfy in \cite{ParSht2} (for example, the coefficients can be smooth periodic, or generic quasi-periodic functions). In particular, periodic magnetic Schr\"odinger operators are covered by our results.

(iii) If $B$ is pseudo-differential, we assume that it is a classical pseudo-differential operator, or, more generally, the operator of classical type. By the latter we mean that the symbol of $B$ admits an asymptotic decomposition in powers of $|\bxi|$ when $|\bxi|\to\infty$; however, these powers do not have to be integer.
Note that operators with the relativistic kinetic energy $\sqrt{(-i\nabla+ \mathbf{A})^2+ m^2}$ are admissible for (almost-)periodic smooth $\mathbf{A}$ and $m\geqslant 0$.

Under these assumptions we prove that the integrated density of states $N(\la)$ has the complete asymptotic expansion \eqref{eq:main_thm1}. This expansion contains powers of $\lambda$ and powers of $\ln\la$; the values of the exponents in the powers of $\la$ depend on the form of $B$, whereas logarithms are raised to integer powers smaller than $d$. Sometimes (as in the case of the magnetic Schr\"odinger operator) we can guarantee that the logarithmic terms are absent (i.e., the corresponding coefficients are zero).

\ber
The main reason why we need assumption (iii) is to match asymptotic expansions in different intervals $I_n$ in Section \ref{reduction section}. If we did not have assumption (iii), we would have obtained the asymptotic expansions containing the general `phase volumes' (like in \cite{Sob}), and it is not clear how to relate the expansions obtained in different intervals $I_n$.
\enr

One immediate and slightly unexpected corollary of \eqref{eq:main_thm1} is as follows:
\bec
Suppose, $H =(-\Delta)^w +B$ with $B$ being {\it periodic} and either differential, or pseudo-differential operator of classical type. Then for sufficiently large $\la$ the spectrum of $H$ is purely absolutely continuous.
\enc
\bep
Since $H$ is periodic, the general Floquet-Bloch theory implies that the spectrum of $H$ is absolutely continuous with the possible exception of eigenvalues of finite multiplicity. If $\la$ is such an eigenvalue, the integrated density of states has a jump at least $|\Gamma^{\dagger}|$ at $\la$, where  $\Gamma^{\dagger}$ is the lattice dual to the lattice of periods of $H$. Due to \eqref{eq:main_thm1}, this cannot happen for large $\la$.
\enp

The approach of our paper is similar to the one of \cite{ParSht2}. In particular, we use the method of gauge transform developed in \cite{Sob}, \cite{Sob1}, and \cite{ParSob}. Nevertheless, there are plenty of new (mostly technical, but sometimes ideological) difficulties arising because the operator $B$ is no longer bounded and no longer local. One example of the new methods employed in this paper is the proof of Lemma \ref{key_lemma}: not only this proof works for unbounded $B$, it also makes Condition D from  \cite{ParSht2} redundant. The biggest increase in technical difficulties comes in Section \ref{contribution section} where we express the contribution to the density of states from various regions in the momentum space as certain complicated integrals and then try to compute these integrals. As a result, our paper is technically more complicated than \cite{ParSht2} (which already was quite difficult to read). Thus, we have reluctantly abandoned the idea of making our paper completely self-contained; we will skip all parts of the argument which are identical (or close) to corresponding parts of \cite{ParSht2} and refer the reader to that paper.
Nevertheless, we will present all the definitions and properties of the important objects.

\ber\label{no proofs remark}
Throughout the article we employ the convention that, if some statement is given without a proof, then an analogous statement can be found in \cite{ParSht2}, and the proof is the same up to obvious modifications.
It comes without saying that the reader is strongly encouraged to read the article \cite{ParSht2} first, before attempting to read this paper.
\enr

{\bf Aknowlegements.}
SM and LP were partially supported by the EPSRC grant EP/ F029721/1. SM was also supported by the Lundbeck Foundation and the European Research Council under the European Community's Seventh Framework Program (FP7/2007--2013)/ERC grant agreement 202859. RS was partially supported by the NSF grant DMS-0901015. The authors would like to thank Gerassimos Barbatis for participation in preliminary discussions which led to this paper. SM would like to express his thanks for hospitality to the University of Athens and ESI Vienna, where part of this work was made.

\section{Preliminaries}\label{introduction section}

For $w> 0$ we consider the operator
\bee\label{eq:Sch}
H =(-\Delta)^w+ B
\ene
acting in $\plainL2(\R^d)$. The action of the pseudo-differential operator $B$ on functions from the Schwarz class $\plainS(\R^d)$ is defined by the formula
\begin{equation*}
(Bf)(\bx):= (2\pi)^{-d/2} \int  b(\bx, \bxi)e^{i\bxi \bx} (\mathcal Ff)(\bxi) d\bxi.
\end{equation*}
Here $\CF$ is the Fourier transform
\begin{equation*}
(\mathcal F f)(\bxi):= (2\pi)^{-d/2} \int e^{-i\bxi\bx}f(\bx) d\bx,\qquad \bxi\in\R^d,
\end{equation*}
the integration is over $\R^d$, and $b$ is the symbol of $B$. We assume that $b(\bx, \bxi)$, $\bx, \bxi\in\R^d$, is a smooth
almost-periodic in $\bx$ complex-valued function and, moreover, that for some countable set $\Bth$ of frequencies we have
\begin{equation}\label{eq:sumf1}
b(\bx, \bxi) = \sum\limits_{\bth\in\Bth}\hat{b}(\bth, \bxi)\be_{\bth}(\bx)
\end{equation}
where
\bee\label{e_theta}
\be_{\bth}(\bx):=e^{i\bth\bx},
\ene
and
\bees
\hat{b}(\bth, \bxi):=\BM_\bx\big(b(\bx,\bxi)\be_{-\bth}(\bx)\big)
\enes
are the Fourier coefficients of $b$ (here $\BM_\bx$ is the mean of an almost-periodic function of $\bx$). We assume that the series \eqref{eq:sumf1} converges absolutely, and that $b$ satisfies the symmetry condition
\begin{equation*}%\label{selfadj:eq}
\hat b(\bth, \bxi) = \overline{\hat b(-\bth, \bxi+\bth)},
\end{equation*}
so that the operator $B$ is formally self-adjoint.
For $R>0$ let $\Id_{\CB_R}$ be the indicator function of the ball $\CB_R:= \big\{\bxi: |\bxi|< R\big\}$. We assume that there exists a constant $C_0$ such that
\bees%\label{bounded inside}
\|b\Id_{\CB_{C_0}}\|_{L_\infty(\R^d\times\R^d)}< \infty,
\enes
and that
\begin{equation}\label{symbol series}
\big(1- \Id_{\CB_{C_0}}(\bxi)\big)b(\bx, \bxi)= \sum_{\iota\in J}|\bxi|^\iota b_\iota\big(\bx, \bxi/|\bxi|\big),
\end{equation}
where $J$ is a discrete subset of $(-\infty, \vark]$ with
\bee\label{condition on kappa}
0\leqslant \vark <2w
\ene
(the first inequality here is assumed for convenience without loss of generality),
and $b_\iota(\bx, \boldeta)$ are smooth functions on $\R^d\times\mathbb S^{d- 1}$ almost-periodic with respect to $\bx$.

Let
\begin{equation}\label{tilde w}
 \tilde w :=(w+ \vark)/2.
\end{equation}
We introduce $\chi\in \plainC\infty(\R_+)$ so that
\begin{equation}\label{tau function}
\chi(r)= \begin{cases}r, & r\geqslant C_0,\\ 0, &r\leqslant C_0/2.\end{cases}
\end{equation}

\ber
Increasing $C_0$ if necessary, we can guarantee that for any $\widetilde J\subset J$ and any $\widetilde\Bth \subset\Bth$ the operator $\widetilde B$ with the symbol $\tilde b$ given by
\begin{equation}\label{tilde b}
\tilde b(\bx, \bxi):= \sum_{\iota\in \widetilde J}\Big(\chi\big(|\bxi|\big)\Big)^\iota\sum_{\bth\in \widetilde\Bth}\hat b_\iota\big(\bth, \bxi/|\bxi|\big)\be_{\bth}(\bx)
\end{equation}
satisfies
\begin{equation}\label{tilde B estimate}
 (-\Delta)^{\tilde w}- |\widetilde B|\geqslant 0.
\end{equation}
\enr

We also assume that the coefficients in the expansion
\bee\label{b_iota}
b_\iota(\bx, \boldeta)= \sum_{\bth\in\Bth}\hat{b}_\iota(\bth, \boldeta)\be_{\bth}(\bx), \qquad \bx\in\R^d, \quad \boldeta\in\mathbb S^{d- 1}, \quad \iota\in J
\ene
can be represented by a series
\begin{equation}\label{series for b}
\hat{b}_\iota(\bth, \eta_1, \dots, \eta_d)= \sum_{\tau\in\N_0^d}\hat{b}_\iota^{(\tau)}(\bth)\eta_1^{\tau_1}\cdots\eta_d^{\tau_d}
\end{equation}
which converges absolutely in a ball of radius greater than one of $\Rd$.

Under the above assumptions $H$ is a selfadjoint operator on the Sobolev space $\plainH{2w}(\R^d)$.
We are interested in the asymptotic behaviour of its integrated density of states $N(\la)$ as
the spectral parameter $\la$ tends to infinity.

\begin{defn}
Let $e(\la;\bx,\by)$ be the kernel of the spectral projection of $H$. We define the integrated density of states as
\bees%\label{eq:densityalmost}
N(\la) :=\BM_{\bx}\big(e(\la;\bx,\bx)\big).
\enes
\end{defn}

It was proved in Theorem 4.1 of \cite{Shu} that for differential operators this definition agrees with the traditional one (at least at its continuity points). The following lemma is proved at the end of Section~4 of \cite{ParSht2}.

\bel\label{norms lemma}
\begin{enumerate}
 \item If $A\ge B$, then $N(\la; A)\le N(\la; B)$.
 \item Suppose, $A= a(\bx, D)$ and $U= u(\bx, D)$ are two pseudo-differential operators with almost-periodic coefficients. Let operator $A$ be elliptic self-adjoint and operator $U$ be unitary. Then $N(\la; A)= N(\la; U^{-1}AU)$.
\end{enumerate}
\enl

Without loss of generality we assume that $\Bth$ (recall \eqref{eq:sumf1}) spans $\R^d$, contains $\mathbf 0$ and is symmetric about $\mathbf 0$; we also put
\bee\label{eq:algebraicsum}
\Bth_k :=\Bth +\Bth +\dots +\Bth
\ene
(algebraic sum taken $k$ times) and $\Bth_{\infty}:=\cup_k\Bth_k=Z(\Bth)$, where for a set $S\subset \R^d$ by $Z(S)$ we denote the set of all finite linear combinations of elements in
$S$ with integer coefficients. The set $\Bth_\infty$ is countable and non-discrete (unless $B$ is periodic).
We will need

\medskip
\paragraph{\bf Condition A} {\it Suppose that $\bth_1,\dots,\bth_d\in \Bth_\infty$. Then $Z(\bth_1,\dots,\bth_d)$ is discrete.}
\medskip

It is easy to see that this condition can be reformulated like this:
suppose, $\bth_1,\dots,\bth_d\in \Bth_\infty$.
Then either $\{\bth_j\}$ are linearly independent, or $\sum_{j=1}^d n_j\bth_j=0$, where $n_j\in\Z$ and not all
$n_j$ are zeros. This reformulation shows that Condition A is generic: indeed, if we are choosing frequencies
of $b$ one after the other, then on each step we have to avoid choosing a new frequency from a countable set of
hyperplanes, and this is obviously a generic restriction. Condition A is obviously satisfied for periodic
$B$, but it becomes meaningful if $B$ is quasi-periodic (i.e., if it is a linear combination of finitely many exponentials).

If $\Bth$ and $J$ are finite, Condition A is all we need. If, however, any (or both) of these sets is infinite, we need other conditions which describe, how well $B$ can be approximated by operators with quasi-periodic symbols. In the proof we are going to work with quasi-periodic approximations of $B$, and we need these conditions to make sure that all estimates in the proof are uniform with respect to these approximations.

We introduce
\begin{equation*}%\label{straight b}
\textsf{b}_\iota(\bth):= \sup_{|\boldeta|= 1}\big|\hat b_\iota(\bth, \boldeta)\big|, \quad \bth\in \Bth.
\end{equation*}

\medskip
\paragraph{\bf Condition B}
{\it Let $k$ be a positive integer. Then there exists $R_0\geqslant C_0$ such that for each $\rho> R_0$ there exist a finite symmetric set $\widetilde\Bth\subset\big(\Bth\cap \CB(\rho^{1/k})\big)$ (where $\CB(r)$ is the ball of radius $r$ centered at $0$) and a finite subset $\widetilde J\subset J$ with
\begin{equation}\label{tilde J cardinality}
\card \widetilde J\leqslant \rho^{1/k}
\end{equation}
such that
\bee\label{eq:condB2}
\sum_{(\bth, \iota)\in (\Bth\times J)\setminus (\widetilde\Bth\times \widetilde J)}\big(1+ |\bth|^2\big)^{\vark/4}|R_0|^{\iota- \vark}\textup{\textsf{b}}_\iota(\bth)\leqslant \rho^{- k}.
\ene}

The last condition we need is a version of the Diophantine condition on the frequencies of $b$. First, we need some definitions. We fix a natural number $\tilde k$ (the choice of $\tilde k$ will be determined later by the order of the remainder in the asymptotic expansion) and denote $\widetilde\Bth'_{\tilde k}:= \widetilde\Bth_{\tilde k}\setminus\{0\}$ (see \eqref{eq:algebraicsum} for the notation).
We say that $\GV$ is a quasi-lattice subspace of dimension $m$, if $\GV$ is a linear span of $m$ linearly independent vectors $\bth_1,\dots,\bth_m$ from $\widetilde\Bth_{\tilde k}$. Obviously, the zero space (which we will denote by $\GX$) is a quasi-lattice subspace of dimension $0$, and $\R^d$ is a quasi-lattice subspace of dimension $d$.
We denote by $\CV_m$ the collection of all quasi-lattice subspaces of dimension $m$ and put $\CV:=\cup_m\CV_m$.
If $\bxi\in\R^d$ and $\GV$ is a linear subspace of $\R^d$, we denote by $\bxi_{\GV}$ the orthogonal projection of $\bxi$ onto $\GV$, and put $\GV^\perp$ to be an orthogonal complement of $\GV$, so that $\bxi= \bxi_{\GV}+ \bxi_{\GV^\perp}$.
Let $\GV,\GU\in\CV$. We say that these subspaces are {\it strongly distinct}, if neither of them is a subspace of the other one. This condition is equivalent to stating that if we put $\GW:=\GV\cap\GU$, then $\dim \GW$ is strictly less than dimensions of $\GV$ and $\GU$. We put $\phi= \phi(\GV, \GU)\in [0, \pi/2]$ to be the angle between them, i.e. the angle between $\GV\ominus\GW$ and $\GU\ominus\GW$, where $\GV\ominus\GW$
is the orthogonal complement of $\GW$ in $\GV$. This angle is non-zero iff $\GV$ and $\GW$ are strongly distinct. We put $s= s(\rho)= s(\widetilde\Bth_{\tilde k}):= \inf\sin\big(\phi(\GV,\GU)\big)$, where the infimum is over all strongly distinct pairs of subspaces from $\CV$,
$R= R(\rho):= \sup_{\bth\in\widetilde\Bth_{\tilde k}}|\bth|$, and $r= r(\rho):= \inf_{\bth\in\widetilde\Bth'_{\tilde k}}|\bth|$. Obviously,
\begin{equation}\label{R(rho)}
R(\rho)= O(\rho^{1/k}),
\end{equation}
where the implied constant can depend on $k$ and $\tilde k$.

\medskip
\paragraph{\bf Condition C} {\it For each fixed $k$ and $\tilde k$ the sets $\widetilde\Bth_{\tilde k}$ can be chosen in such a way that for sufficiently large $\rho$ the number of elements in $\widetilde\Bth_{\tilde k}$ satisfies $\card\widetilde\Bth_{\tilde k}\le\rho^{1/k}$ and we have
\bee\label{eq:condC1}
s(\rho)\ge\rho^{-1/k}
\ene
and
\bees%\label{eq:condC2}
r(\rho)\ge\rho^{-1/k},
\enes
where the implied constant (i.e. how large should $\rho$ be) can depend on $k$ and $\tilde k$.}
\medskip

\ber\label{rem:condB}
Note that Condition C is automatically satisfied for quasi-periodic and smooth periodic $B$; see \cite{ParSht2} for further
discussion of this condition.
\enr

Condition A implies the following statement, which will be used crucially in our constructions.

\begin{cor}\label{D type corollary}
Suppose, $\bth_1, \dots, \bth_{l}\in \widetilde\Bth_{\tilde k}$, $l\leqslant d- 1$. Let $\GV$ be the span of $\bth_1, \dots, \bth_{l}$. Then each element of the set $\widetilde\Bth_{\tilde k}\cap\GV$ is a linear combination of $\bth_1, \dots, \bth_{l}$ with rational coefficients. Since the set $\widetilde\Bth_{\tilde k}\cap\GV$ is finite, this implies that the set $Z(\widetilde\Bth_{\tilde k}\cap\GV)$ is discrete and is, therefore, a lattice in $\GV$.
\end{cor}

From now on, we always assume that $B$ satisfies all the conditions from this section; we will also denote
\begin{equation*}
\rho:= \la^{1/2w}.
\end{equation*}
Now we can formulate our main theorem.

\bet\label{main_thm}
Let $H$ be an operator \eqref{eq:Sch} satisfying Conditions {\rm A, B} and {\rm C}.
Then for each $K\in \R$ there exists a finite positive integer $L$ and a finite subset $J_0\subset J$ such that
\bee\label{eq:main_thm1}
\begin{split}
&N(\rho^{2w})\\ &= \sum_{q= 0}^{d- 1}\sum_{h= 0}^{L}\sum_{\iota_1, \dots, \iota_h\in J_0}\sum_{j= 0}^{[K+ d+ (2- 2w)h+ \iota_1+ \cdots+ \iota_h]}C_{q\, h\, j}^{\iota_1\cdots \iota_h}\rho^{d+ (2- 2w)h+ \iota_1+ \cdots+ \iota_h- j}\ln^q\rho+ O(\rho^{-K}).
\end{split}
\ene
as $\rho\to\infty$.
\ent
\ber\label{spurious remark}
The powers of $\rho$ present in \eqref{eq:main_thm1} are equal to $d+ (2- 2w)h+ \iota_1+ \cdots+ \iota_h- j$, and the first impression is that there are far too many of them (indeed, a priori the set of all such powers can be dense in $\R$, for instance). However, many of these powers are, in fact, spurious (i.e. the corresponding coefficients
$C_{q\, h\, j}^{\iota_1\cdots \iota_h}$ are zero). This happens, for example, when $d+ (2- 2w)h+ \iota_1+ \cdots+ \iota_h- j>d$ (for obvious reasons). Equally obviously, these powers do not `multiply' when we increase $K$. This means that if $K_1<K_2$, then expansion \eqref{eq:main_thm1} with $K=K_2$ does not contain extra terms with $d+ (2- 2w)h+ \iota_1+ \cdots+ \iota_h- j>-K_1$, compared to this expansion for $K=K_1$.
\enr

In the case of magnetic Schr\"odinger operators, Theorem \ref{main_thm} and calculations similar to those of \cite{HitPol} and \cite{ParSht2} imply that most of the terms in \eqref{eq:main_thm1}
will indeed disappear:
\bec
For each $K\in\N$ we have:
\bee\label{eq:main_cor1}
N(\la)=\la^{d/2}\bigg(C_d+\sum\limits_{j=1}^{K}e_j\la^{-j}+o(\la^{-K})\bigg)
\ene
as $\la\to\infty$.
\enc

\ber
By taking the Laplace transform of \eqref{eq:main_thm1}, one can obtain an asymptotic expansion of the (regularised) heat trace as $t\to 0$. However, it seems that using the approach of \cite{HitPol} and \cite{HitPol1}, it is possible to obtain even stronger results (the pointwise asymptotic expansion of the heat kernel).
\enr

\ber
Of course, formula \eqref{eq:main_lem1} cannot be differentiated; moreover, we do not even know if in the almost periodic case $N(\lambda)$ is strictly increasing. However, in the periodic Schr\"odinger case there are some results on the high-energy behaviour of the (non-integrated) density of states, see e. g. \cite{MorParPch}.
\enr

Given Conditions B and C, we want to introduce the following definition. We say that a non-negative function $f= f(\rho)= f(\rho; k, \tilde k)$ satisfies the estimate $f(\rho)\le \rho^{0+}$ (resp. $f(\rho)\ge \rho^{0-}$), if for each positive $\ep$ and for each $\tilde k$ we can achieve $f(\rho)\le \rho^{\ep}$ (resp. $f(\rho)\ge \rho^{-\ep}$) for sufficiently large $\rho$ by choosing parameter $k$ from Conditions B and C sufficiently large. For example, we have
\begin{equation}\label{bound on R}
R(\rho)\le\rho^{0+},
\end{equation}
$\card \widetilde\Bth\le\rho^{0+}$, $s(\rho)\ge\rho^{0-}$, and $r(\rho)\ge\rho^{0-}$.

Throughout the paper, we always assume that the value of $k$ is chosen
sufficiently large so that all inequalities of the form $\rho^{0+}\le\rho^{\ep}$ or $\rho^{0-}\ge\rho^{-\ep}$
we encounter in the proof are satisfied.

The next statement proved in \cite{ParSht2} is an example of how this new notation is used.
\bel\label{lem:coefficients}
Suppose, $\bth, \bmu_1,\dots,\bmu_d\in\widetilde\Bth'_{\tilde k}$, the set $\{\bmu_j\}$ is linearly independent, and $\bth=\sum_{j=1}^db_j\bmu_j$. Then each non-zero coefficient
$b_j$ satisfies
\bees%\label{eq:coefficients}
\rho^{0-}\le |b_j| \le \rho^{0+}.
\enes
\enl

In this paper, by $C$ or $c$ we denote positive constants, the exact value of which can be different each time they
occur in the text, possibly even in the same formula. On the other hand, the constants which are labeled (like $C_1$, $c_3$, etc) have their values being fixed throughout the text.
Given two positive functions $f$ and $g$, we say that $f\gtrsim g$, or $g\lesssim f$, or $g=O(f)$ if the ratio $g/f$ is bounded. We say $f\asymp g$ if $f\gtrsim g$ and $f\lesssim g$.

We will also need a number of auxiliary constants. Let us choose numbers $\{\a_j\}_{j= 1}^d$, $\b$, $\vartheta$, and $\varsigma$ satisfying
\bee\label{beta and alphas}
\max\{1- w + \vark/2, 1/2\}< \b< \a_1< \a_2< \cdots < \a_d< \vartheta< \varsigma< 1
\ene
(recall \eqref{condition on kappa}), and set
\bee\label{alpha}
\a:= \vark/\b.
\ene

\section{Reduction to a finite interval of spectral parameter}\label{reduction section}

To begin with, we choose sufficiently large $\rho_0> C_0$ (to be fixed later on) and for $n\in \N$ put $\rho_n:= 2\rho_{n- 1}= 2^n\rho_0$. We also define the intervals $I_n:= [\rho_n, 4\rho_n]$.
The proof of Theorem~\ref{main_thm} will be based on the following lemma:

\bel\label{main_lem}
For each $M\in\R$ there exist $L> 0$ and a finite subset $J_0\subset J$ such that for every $n\in \N$ and $\rho\in I_n$
\bee\label{eq:main_lem1}
N(\rho^{2w})= \sum_{q= 0}^{d- 1}\sum_{h= 0}^{L}\sum_{\iota_1, \dots, \iota_h\in J_0}\sum_{j= 0}^{[\frac{d+ M}{1- \varsigma}]}C_{q\, h\, j}^{\iota_1\cdots \iota_h}(n, M)\rho^{d+ (2- 2w)h+ \iota_1+ \cdots+ \iota_h- j}\ln^q\rho+ O(\rho_n^{-M}).
\ene
Here, $C_{q\, h\, j}^{\iota_1\cdots \iota_h}(n, M)$ are some real numbers satisfying
\bee\label{eq:main_lem2}
C_{q\, h\, j}^{\iota_1\cdots \iota_h}(n, M)= O(\rho_n^{-2\b h+ \varsigma j}).%\ \ c<4/6.
\ene
The constants in the $O$-terms do not depend on $n$ (but they may depend on $M$).
\enl
\ber\label{rem:new1}
Note that \eqref{eq:main_lem1} is not a `proper' asymptotic formula, since the coefficients are allowed to grow with $n$ (and, therefore, with $\rho$).
\enr

Some of the powers of $\rho$ on the right hand side of \eqref{eq:main_lem1} may coincide. In order to avoid the ambiguity let us redefine coefficients $C_{q\, h\, j}^{\iota_1\cdots \iota_h}(n, M)$ in such a way that, for any given values of $q$ and $d+ (2- 2w)h+ \iota_1+ \cdots+ \iota_h- j$, only the coefficient with the minimal possible value of $h$ and maximal possible values of $j$, $\iota_1, \dots, \iota_h$ (in this order) is nonzero. Note that these new coefficients still satisfy \eqref{eq:main_lem2}.

Let us prove Theorem \ref{main_thm} assuming that we have proved Lemma \ref{main_lem}. Let $M$ be fixed.
Denote the sum on the right hand side of \eqref{eq:main_lem1} by $N_n(\rho^{2w})$.
Then, for $n\geqslant 1$, whenever $\rho\in I_{n-1}\cap I_n=[\rho_n,2\rho_n]$, we have:
\begin{equation}\label{difference of Ns}
\begin{split}
&N_n(\rho^{2w})- N_{n- 1}(\rho^{2w})\\ &= \sum_{q= 0}^{d- 1}\sum_{h= 0}^{L}\sum_{\iota_1, \dots, \iota_h\in J_0}\sum_{j= 0}^{[\frac{d+ M}{1- \varsigma}]} t_{q\, h\, j}^{\iota_1\cdots \iota_h}(n, M)\rho^{d+ (2- 2w)h+ \iota_1+ \cdots+ \iota_h- j}\ln^q\rho+ O(\rho_n^{-M}),
\end{split}
\end{equation}
where
\bees
t_{q\, h\, j}^{\iota_1\cdots \iota_h}(n, M):= C_{q\, h\, j}^{\iota_1\cdots \iota_h}(n, M)- C_{q\, h\, j}^{\iota_1\cdots \iota_h}(n- 1, M).
\enes
On the other hand, since for $\rho\in I_{n- 1}\cap I_n$ we have both
$N(\rho^{2w})= N_n(\rho^{2w})+ O(\rho_n^{-M})$ and
$N(\rho^{2w})= N_{n- 1}(\rho^{2w})+ O(\rho_n^{-M})$, this implies that
\begin{equation}\label{sum is O}
\sum_{q= 0}^{d- 1}\sum_{h= 0}^{L}\sum_{\iota_1, \dots, \iota_h\in J_0}\sum_{j= 0}^{[\frac{d+ M}{1- \varsigma}]} t_{q\, h\, j}^{\iota_1\cdots \iota_h}(n, M)\rho^{d+ (2- 2w)h+ \iota_1+ \cdots+ \iota_h- j}\ln^q\rho= O(\rho_n^{-M}).
\end{equation}

\becl For each combination of indices present on the right hand side of \eqref{difference of Ns} we have:
\begin{equation}\label{claim on t}
t_{q\, h\, j}^{\iota_1\cdots \iota_h}(n, M)= O(\rho_n^{j- M- d+ (2w- 2)h- \iota_1- \cdots- \iota_h}\ln^{d- 1- q}\rho_n).
\end{equation}

\encl
\bep
Put $y:= \rho_n/\rho$ and let
\begin{equation}\label{taus}
\tau_{p\, h\, j}^{\iota_1\cdots \iota_h}(n, M):= \rho_n^{M+ d+ (2- 2w)h+ \iota_1+ \cdots+ \iota_h- j}\sum_{q= p}^{d- 1}\binom qp(-1)^p t_{q\, h\, j}^{\iota_1\cdots \iota_h}(n, M)\ln^{q- p}\rho_n.
\end{equation}
Then by \eqref{sum is O} for $y\in [1/2, 1]$
\bee\label{Cramer}
P(y):= \sum_{p= 0}^{d- 1}\sum_{h= 0}^{L}\sum_{\iota_1, \dots, \iota_h\in J_0}\sum_{j= 0}^{[\frac{d+ M}{1- \varsigma}]}\tau_{p\, h\, j}^{\iota_1\cdots \iota_h}(n, M)y^{j- d+ (2w- 2)h- \iota_1- \cdots- \iota_h}\ln^p y= O(1).
\ene
Let us denote by $h_1, \dots, h_T$ the functions $y^{j- d+ (2w- 2)h- \iota_1- \cdots- \iota_h}\ln^p y$ entering the sum in \eqref{Cramer} with non-zero coefficients; these functions are linearly independent on the interval $[1/2, 1]$. Therefore, there exist points $y_1,...,y_{T}\in [1/2, 1]$ such that the determinant of the matrix $\big(h_j(y_l)\big)_{j, l= 1}^{T}$ is non-zero. Now \eqref{Cramer} and the Cramer's Rule imply that the values of $\tau_{p\, h\, j}^{\iota_1\cdots \iota_h}(n, M)$ are fractions with a bounded expression in the numerator and a fixed non-zero number in the denominator. Therefore,
\begin{equation}\label{tau is O(1)}
\tau_{p\, h\, j}^{\iota_1\cdots \iota_h}(n, M)= O(1).
\end{equation}
Thus, choosing $p= d- 1$ in \eqref{taus}, we obtain
\[
t_{d- 1\, h\, j}^{\iota_1\cdots \iota_h}(n, M)= O(\rho_n^{j- M- d+ (2w- 2)h- \iota_1- \cdots- \iota_h}).
\]
Now we can put $p= d- 2$ into \eqref{tau is O(1)} and obtain
\[
t_{d- 1\, h\, j}^{\iota_1\cdots \iota_h}(n, M)= O(\rho_n^{j- M- d+ (2w- 2)h- \iota_1- \cdots- \iota_h}\ln\rho_n).
\]
Continuing this process until $p= 0$, we obtain \eqref{claim on t}.
\enp

Thus, for $j< M+ d+ (2- 2w)h+ \iota_1+ \cdots+ \iota_h$, the series $\sum_{m=0}^\infty t_{q\, h\, j}^{\iota_1\cdots \iota_h}(m, M)$ is absolutely convergent; moreover, for such $j$ we have:
\bees
\begin{split}
& C_{q\, h\, j}^{\iota_1\cdots \iota_h}(n, M)= C_{q\, h\, j}^{\iota_1\cdots \iota_h}(0, M)+ \sum_{m= 1}^n
t_{q\, h\, j}^{\iota_1\cdots \iota_h}(m, M)\\ &= C_{q\, h\, j}^{\iota_1\cdots \iota_h}(0, M)+ \sum_{m= 1}^\infty t_{q\, h\, j}^{\iota_1\cdots \iota_h}(m, M)+ O(\rho_n^{j- M- d+ (2w- 2)h- \iota_1- \cdots- \iota_h}\ln^{d- 1- q}\rho_n)\\ & =: C_{q\, h\, j}^{\iota_1\cdots \iota_h}(M)+ O(\rho_n^{j- M- d+ (2w- 2)h- \iota_1- \cdots- \iota_h}\ln^{d- 1- q}\rho_n),
\end{split}
\enes
where we have denoted $C_{q\, h\, j}^{\iota_1\cdots \iota_h}(M):= C_{q\, h\, j}^{\iota_1\cdots \iota_h}(0, M)+ \sum_{m =1}^\infty t_{q\, h\, j}^{\iota_1\cdots \iota_h}(m, M)$. For bigger values of $j$ we use \eqref{eq:main_lem2} and \eqref{beta and alphas} to obtain
\bees
\begin{split}\label{intermediate js} &\sum_{q= 0}^{d- 1}\sum_{h= 0}^{L}\sum_{\iota_1, \dots, \iota_h\in J_0}\sum_{\substack{j\geqslant M+ d+ (2- 2w)h+ \iota_1+ \cdots+
\iota_h}}^{[\frac{d+ M}{1- \varsigma}]} \big|C_{q\, h\, j}^{\iota_1\cdots \iota_h}(n, M)\big|\rho^{d+ (2- 2w)h+ \iota_1+ \cdots+ \iota_h- j}\ln^q\rho\\ &\lesssim\sum_{q= 0}^{d- 1} \sum_{h= 0}^{L}\sum_{\iota_1, \dots, \iota_h\in J_0}\rho_n^{\varsigma d+ (2\varsigma- 2\beta- 2\varsigma w+ \varsigma\vark)h- (1- \varsigma)M}\ln^q\rho_n\lesssim \rho_n^{\varsigma d- (1- \varsigma)M}\ln^{d- 1}\rho_n.
\end{split}
\enes

Thus, when $\rho\in I_n$, we have:
\bees%\label{eq:main_lem11}
\begin{split}
N(\rho^{2w})&= \sum_{q= 0}^{d- 1}\sum_{h= 0}^{L}\sum_{\iota_1,
\dots, \iota_h\in J_0}\sum_{j= 0}^{[M+ d+ (2- 2w)h+ \iota_1+ \cdots+
\iota_h]} C_{q\, h\, j}^{\iota_1\cdots \iota_h}(M)\rho^{d+ (2- 2w)h+
\iota_1+ \cdots+ \iota_h- j}\ln^q\rho\\ &+ O(\rho^{-M}\ln^{d-
1}\rho)+ O(\rho^{\varsigma d- (1- \varsigma)M}\ln^{d- 1}\rho).
\end{split}
\enes
Since the constants in $O$ terms do not depend on $n$, it is sufficient to choose
\[
M:= \big[(\varsigma d+ K)/(1- \varsigma)\big]+ 1
\]
to get \eqref{eq:main_thm1} for all $\rho\geqslant \rho_0$.

The rest of the paper is devoted to proving Lemma \ref{main_lem}.
The first step of the proof is fixing $n$ and fixing large $\tilde k$ and $k$. The precise value of $\tilde k$ will be chosen later; the only restriction on it will be to satisfy inequality \eqref{eq:kM} (it says that
the more asymptotic terms we want to have in \eqref{eq:main_lem1}, the bigger $\tilde k$ we need to choose; note that
the choice of $\tilde k$ does not depend on $k$).
We will have several requirements on how large $k$ should be (most of them will be of the form $\rho_n^{0+}< \rho_n^{\ep}$
or $\rho_n^{0-}>\rho_n^{-\ep}$); each time we have such an inequality, we assume that $k$ is chosen sufficiently
large to satisfy it.
\ber\label{k remark}
Our choice of $k$ will only depend on $M$, $w$, $\vark$, and the constants introduced in \eqref{beta and alphas}. The set $J_0$ in Lemma~\ref{main_lem} can be chosen to be
\begin{equation}\label{tilde J condition}
J_0:= J\cap [\vark -d -M -1, \vark].
\end{equation}
\enr

The first requirement on $k$ we have is that
\begin{equation}\label{k fist condition}
k> d+ M+ \vark(d+ M)/(w- \vark) -2w.
\end{equation}
After fixing $\tilde k$ and $k$ we get $R_0$ from Condition B. Then, taking 
\begin{equation}\label{rho_0 condition}
\rho_0\geqslant R_0
\end{equation}
and fixing $n$, we choose $\widetilde\Bth$ and $\widetilde J$ so that Conditions B and C are satisfied for $\rho:= 4\rho_n$. Without loss of generality we may assume that $\widetilde J \supset J_0$.
Then we introduce an auxiliary pseudo--differential operator $\widetilde B$ with the symbol $\tilde b$ given by \eqref{tilde b}.

From now on we prove Lemma~\ref{main_lem} for $B= \widetilde B$ and with $J_0$ replaced by $\widetilde J$. However, in view of \eqref{tilde J cardinality} and \eqref{beta and alphas}, the results with $\widetilde J$ and $J_0$ are equivalent. Afterwards, in Section~\ref{final section} we will prove that the asymptotics \eqref{eq:main_lem1} for the original $B$ follows from Condition B and \eqref{eq:main_lem1} for $\widetilde B$.

\section{Pseudo-differential operators}\label{PsDO section}

Most of the material in this and several subsequent sections is very
similar to the corresponding sections of \cite{ParSht2} and
\cite{ParSob}, as are the proofs of most of the statements.
Therefore, we will often omit the proofs, instead referring the
reader to \cite{ParSht2}, \cite{Sob}, and \cite{ParSob}.

\subsection{Classes of PDO's}\label{classes:subsect}
Before we define the pseudo-differential operators (PDO's), we introduce the relevant classes of symbols.
Let $b = b(\bx, \bxi)$, $\bx, \bxi\in\R^d$, be an almost-periodic (in $\bx$) complex-valued function and, moreover, for some countable set $\hat{\Bth}$ of frequencies (we always assume that $\hat\Bth$ is symmetric and contains $0$; starting from the middle of this section, $\hat\Bth$ will be assumed to be finite)
\begin{equation}\label{eq:sumf}
b(\bx, \bxi) = \sum\limits_{\bth\in\hat{\Bth}}\hat{b}(\bth, \bxi)\be_{\bth}(\bx),
\end{equation}
where
\bees
\hat{b}(\bth, \bxi):=\BM_\bx\big(b(\bx,\bxi)\be_{-\bth}(\bx)\big)
\enes
are the Fourier coefficients of $b(\cdot, \bxi)$ (recall that $\BM$ is the mean of an almost-periodic function). We always assume that \eqref{eq:sumf} converges absolutely.
Let us now define the classes of symbols we will consider and operators associated with them.
For $\bxi\in\R^d$ let $\lu \bxi \ru := \sqrt{1+|\bxi|^2}$. We notice that
\begin{equation}\label{weight:eq}
\lu \bxi + \boldeta\ru\le 2\lu\bxi\ru \lu\boldeta\ru, \ \forall
\bxi, \boldeta\in \R^d.
\end{equation}
We say that a symbol $b$ belongs to the class $\BS_{\a}= \BS_{\a}(\beta)= \BS_{\a}(\beta, \hat{\Bth})$, if for any $l\ge 0$ and any non-negative $s\in\Z$ the conditions
\begin{equation}\label{1b1:eq}
\1 b \1^{(\a)}_{l, s}:= \max_{|\bs| \le s}\sum\limits_{\bth\in\hat{\Bth}}\lu \bth\ru^{l}\sup_{\bxi}\lu\bxi\ru^{(-\a+ |\bs|)\beta}\big|\BD_{\bxi}^{\bs}\hat b(\bth, \bxi)\big|< \infty, \quad |\bs|= s_1+ s_2+ \dots+ s_d,
\end{equation}
are fulfilled.
The quantities \eqref{1b1:eq} define norms on the class $\BS_\a$. Note that $\BS_\a$ is an increasing function of $\a$,
i.e. $\BS_{\a}\subset\BS_{\g}$ for $\a < \g$.

Given $\bth\in \Rd$, let us introduce a linear map $\nabla_{\bth}$ on symbols which acts according to the rule
\begin{equation}\label{Delta}
\widehat{(\nabla_{\bth} a)}(\bphi, \bxi):= \hat a(\bphi, \bxi+ \bth)- \hat a(\bphi, \bxi).
\end{equation}
If the Fourier transform of the symbol is factorized, i.e.
\[
\hat a(\bphi, \bxi)= \prod_{q= 1}^Q\hat a_q(\bphi, \bxi),
\]
then the action of $\nabla_{\bth}$ can be written as a sum of actions on each factor separately:
\begin{equation}\label{nabla of product}
\widehat{(\nabla_{\bth} a)}(\bphi, \bxi)= \sum_{q= 1}^Q\prod_{l= 1}^{q- 1}\hat a_l(\bphi, \bxi+ \bth)\widehat{(\nabla_{\bth} a_q)}(\bphi, \bxi)\prod_{s= q- 1}^Q\hat a_s(\bphi, \bxi).
\end{equation}

For later reference we mention here the following convenient bound that follows from definition
\eqref{1b1:eq} and property \eqref{weight:eq}:
\bee
\sum\limits_{\bth\in\hat{\Bth}}\lu\bth\ru^{l}\sup_{\bxi}\, \lu\bxi\ru^{(-\a+ s+ 1)\b}\Big(\big|\BD^{\bs}_{\bxi}\widehat{(\nabla_{\boldeta} b)}(\bth, \bxi)\big|\Big)\le C\1 b\1^{(\a)}_{l, s+ 1}  \lu\boldeta\ru^{|\a- s- 1|\b}
|\boldeta|, \ s= |\bs|, \label{differ:eq}
\ene
with a constant $C$ depending only on $\a, s$, and $\beta$.
The estimate \eqref{differ:eq} implies that for all $\boldeta$ with $|\boldeta|\le C$ we have a uniform bound
\begin{equation*}%\label{differ1:eq}
\1 \nabla_{\boldeta} b\1^{(\a-1)}_{l, s}\le C \1 b\1^{(\a)}_{l, s+1}|\boldeta|.
\end{equation*}

Now we define the PDO $\op(b)$ in the usual way:
\begin{equation}\label{eq:deff}
\op(b)u(\bx) = (2\pi)^{-d/2} \int  b(\bx, \bxi)
e^{i\bxi \bx} (\mathcal Fu)(\bxi) d\bxi,
\end{equation}
the integral being over $\Rd$. Under the condition $b\in\BS_\a$ the integral on the r.h.s. is clearly finite for any $u$ from the Schwarz class $\plainS(\Rd)$. Moreover, the property $b\in \BS_0$ guarantees the boundedness of $\op(b)$ in $\plainL2(\Rd)$, see Proposition~\ref{bound:prop}. Unless otherwise stated, from now on $\plainS(\Rd)$ is taken as a natural domain for all PDO's when they act in $\plainL2(\Rd)$.

Applying the standard regularization procedures to definition \eqref{eq:deff} (see, e.g., \cite{Shu0}), we can also consider the action of $\op(b)$ on the exponentials $\be_{\bnu}$, $\bnu\in\R^d$. Namely, we have
\bees%\label{eq:actionbe}
\op(b)\be_{\bnu}=\sum_{\bth\in\hat\Bth}\hat{b}(\bth, \bnu)\be_{\bnu+\bth}.
\enes
This action can be extended by linearity to all quasi-periodic functions (i.e. finite linear combinations of $\be_{\bnu}$ with different $\bnu$).
By taking the closure, we can extend this action of $\op(b)$ to the Besicovitch space $\plainB_2(\R^d)$. This is the space of all formal sums
\bees
\sum_{j=1}^\infty a_{j}\be_{\bth_j}(\bx), \quad\textrm{with}\quad \sum_{j=1}^\infty |a_{j}|^2<+\infty.
\enes
It is known (see \cite{Shu0}) that the spectra of $\op(b)$ acting in $\plainL2(\R^d)$ and $\plainB_2(\R^d)$ are the same, although the types of
the spectra can be entirely different. It is very convenient, when working with the gauge transform constructions, to assume that all the operators involved act in $\plainB_2(\R^d)$, although in the end we will return to operators acting in $\plainL2(\R^d)$. This trick (working with operators acting in $\plainB_2(\R^d)$) is similar to working with fibre operators in the periodic case in the sense that we can freely consider the action of an operator on one, or finitely many, exponentials \eqref{e_theta}, despite the fact that these exponentials do not belong to our original function space.

Moreover, if the order $\al=0$ then by continuity this action can be extended to all of $\plainB_2(\R^d)$, and the extension has the same norm as $\op(b)$ acting in $\plainL2$ (see \cite{Shu0}). Thus, in what follows, when we speak about a pseudo-differential operator with almost-periodic symbol acting in $\plainB_2$, we mean that its domain is either whole $\plainB_2$ (when the order is non-positive), or the space of all quasi-periodic functions (for operators with positive order). And, when we make a statement about the norm of a pseudo-differential operator with almost-periodic symbol, we will not specify whether the operator acts in $\plainL2(\R^d)$ or  $\plainB_2(\R^d)$, since these norms are the same.

\subsection{Some basic results on the calculus of almost-periodic PDO's}
We begin by listing some elementary results for almost-periodic PDO's. The proofs are very similar (with obvious changes) to the proof of analogous statements in \cite{Sob}.

\begin{prop}\label{bound:prop}
Suppose that $b$ satisfies \eqref{eq:sumf} and that $\1 b\1^{(0)}_{0, 0}<\infty$. Then $\op(b)$ is bounded in both $\plainL2(\R^d)$ and $\plainB_2(\R^d)$ and $\big\|\op(b)\big\|\le \1 b \1^{(0)}_{0, 0}$.
\end{prop}

In what follows, {\it if we need to calculate a product of two (or more) operators with some symbols $b_j\in\BS_{\a_j}(\hat{\Bth}_j)$ we will always consider that $b_j\in\BS_{\a_j}(\sum_j\hat{\Bth}_j)$ where, of course, all extra terms are assumed to have zero coefficients in front of them}.

Since $\op(b) u\in\plainS(\Rd)$ for any $b\in\BS_{\a}$ and $u\in
\plainS(\Rd)$, the product $\op(b) \op(g)$, $b\in \BS_{\a}(\hat{\Bth}_1), g\in
\BS_{\g}(\hat{\Bth}_2)$, is well defined on $\plainS(\Rd)$. A straightforward
calculation leads to the following formula for the symbol $b\circ g
$ of the product $\op(b)\op(g)$:
\begin{equation*}
(b\circ g)(\bx, \bxi) = \sum_{\bth\in\hat{\Bth}_1,\, \bphi\in\hat{\Bth}_2}
\hat b(\bth, \bxi +\bphi) \hat g(\bphi, \bxi) e^{i(\bth+\bphi)\bx},
\end{equation*}
and hence
\begin{equation}\label{prodsymb:eq}
\widehat{(b\circ g)}(\bchi, \bxi) =
\sum_{\bth +\bphi = \bchi} \hat b (\bth, \bxi +\bphi) \hat g(\bphi,
\bxi),\ \bchi\in\hat{\Bth}_1+\hat{\Bth}_2,\ \bxi\in \Rd.
\end{equation}

We have
\begin{prop}\label{product:prop}
Let $b\in\BS_{\a}(\hat{\Bth}_1)$,\ $g\in\BS_{\g}(\hat{\Bth}_2)$. Then $b\circ g\in\BS_{\a+\g}(\hat{\Bth}_1+\hat{\Bth}_2)$
and
\begin{equation*}
\1 b\circ g\1^{(\a+\g)}_{l,s} \le C \1 b\1^{(\a)}_{l,s} \1 g\1^{(\g)}_{l+(|\a|+s)\beta,s},
\end{equation*}
with the constant $C$ depending only on $l$, $\alpha$, and $s$.
\end{prop}

We are also interested in the estimates for symbols of commutators.
For PDO's $A, \Psi_l, \ l = 1, 2, \dots ,N$, denote
\begin{gather*}
\ad(A; \Psi_1, \Psi_2, \dots, \Psi_N):= i\bigl[\ad(A; \Psi_1, \Psi_2, \dots, \Psi_{N-1}), \Psi_N\bigr],\\
\ad(A; \Psi):= i[A, \Psi],\quad \ad^N(A; \Psi):= \ad(A; \Psi, \Psi, \dots, \Psi),\quad \ad^0(A; \Psi):= A.
\end{gather*}
For the sake of convenience, we use the notation $\ad(a;  \psi_1, \psi_2, \dots, \psi_N)$ and $\ad^N(a, \psi)$ for the symbols of
multiple commutators.

Let
\[
\supp\hat b:= \big\{\bth\in\Rd: \hat b(\bth, \cdot)\not\equiv 0\big\}.
\]
It follows from \eqref{prodsymb:eq} that the Fourier coefficients of the symbol $\ad(b,g)$ are given by
\begin{equation}\label{comm:eq}
\widehat{\ad(b, g)}(\bchi, \bxi)= i\!\!\!\sum_{\bth\in (\supp\hat b)\cup(\bchi- \supp\hat g)}\!\!\!\bigl[\widehat{(\nabla_{\bchi- \bth} b)}(\bth, \bxi)\hat g(\bchi- \bth, \bxi)- \hat b(\bth, \bxi)\widehat{(\nabla_{\bth}g)}(\bchi- \bth, \bxi)\bigr].
\end{equation}

\begin{prop}\label{commut0:prop}
Let $b\in \BS_{\a}(\hat{\Bth})$ and $g_j\in\BS_{\g_j}(\hat{\Bth}_j)$,\ $j = 1, 2, \dots, N$. Then\\ $\ad(b; g_1, \dots, g_N) \in\BS_{\g}(\hat{\Bth}+\sum_j\hat{\Bth}_j)$ with
$$
\g = \a+\sum_{j=1}^N(\g_j-1),
$$
and
\begin{equation*}%\label{commutator:eq}
\1 \ad(b; g_1, \dots, g_N)\1^{(\g)}_{l,s} \le C \1 b\1^{(\a)}_{p,s +N}
\prod_{j =1}^N \1 g_j\1^{(\g_j)}_{p, s +N -j +1},
\end{equation*}
where $C$ and $p$ depend on $l, s, N, \a$ and $\g_j$.
\end{prop}

\section{Resonant regions}

We now define resonant regions and mention some of their properties. This material is essentially identical to Section~5 of \cite{ParSht2}, where the reader can find the proofs of all the statements of this section.

Recall the definition of the set $\Bth= \widetilde\Bth$ as well as of the quasi-lattice subspaces from Section~\ref{introduction section}. As before, by $\Bth_{\tilde k}$ we denote the algebraic sum of $\tilde k$ copies of $\Bth$; remember that we consider $\tilde k$ fixed. We also put $\Bth'_{\tilde k}:=\Bth_{\tilde k}\setminus\{0\}$. For each $\GV\in\CV$ we put $S_{\GV}:= \big\{\bxi\in\GV,\ |\bxi|=1\big\}$.
For each non-zero $\bth\in\R^d$ we put $\bn(\bth):=\bth|\bth|^{-1}$.

Let $\GV\in\CV_m$. We say that $\GF$ is a {\it flag} generated by $\GV$, if $\GF$ is a sequence $\GV_j\in\CV_j$ ($j= 0, 1, \dots, m$) such that $\GV_{j- 1}\subset\GV_j$ and $\GV_m= \GV$. We say that $\{\bnu_j\}_{j= 1}^m$ is a sequence generated by $\GF$ if $\bnu_j\in\GV_j\ominus\GV_{j- 1}$ and $\|\bnu_j\|= 1$ (obviously, this condition determines each $\bnu_j$ up to multiplication by $-1$). We denote by $\CF(\GV)$ the collection of all flags generated by $\GV$. We put
\begin{equation}\label{L_j}
L_j:= \rho_n^{\al_j},
\end{equation}
recall \eqref{beta and alphas}.

Let $\bth\in\Bth'_{\tilde k}$. The {\it resonant region} generated by $\bth$ is defined as
\bee\label{eq:1}
\Lambda(\bth):= \Big\{\bxi\in\R^d,\ \big|\lu\bxi, \bn(\bth)\ru\big|\le L_1\Big\}.
\ene
Suppose, $\GF\in\CF(\GV)$ is a flag and $\{\bnu_j\}_{j= 1}^m$ is a sequence generated by $\GF$. We define
\bee\label{eq:2}
\Lambda(\GF):= \Big\{\bxi\in\R^d,\ \big|\lu\bxi,\bnu_j\ru\big|\le L_j\Big\}.
\ene
If $\dim\GV= 1$, definition \eqref{eq:2} is reduced to \eqref{eq:1}.
Obviously, if $\GF_1\subset\GF_2$, then $\Lambda(\GF_2)\subset\Lambda(\GF_1)$.

Suppose, $\GV\in\CV_j$. We denote
\bees%\label{eq:Bxi1}
\Bxi_1(\GV) :=\cup_{\GF\in\CF(\GV)}\Lambda(\GF).
\enes
Note that $\Bxi_1(\GX)= \R^d$ and $\Bxi_1(\GV)= \Lambda(\bth)$ if $\GV\in\CV_1$ is spanned by $\bth$.
Finally, we put
\bee\label{eq:Bxi}
\Bxi(\GV):= \Bxi_1(\GV)\setminus\big(\cup_{\GU\supsetneq\GV}\Bxi_1(\GU)\big)= \Bxi_1(\GV)
\setminus\big(\cup_{\GU\supsetneq\GV}\cup_{\GF\in\CF(\GU)}\Lambda(\GF)\big).
\ene
We call $\Bxi(\GV)$ the resonance region generated by $\GV$.
Very often, the region $\Bxi(\GX)$ is called the non-resonance region. We, however, will omit using
this terminology since we will treat all regions $\Bxi(\GV)$ in the same way.

The first set of properties follows immediately from the definitions.

\bel\label{lem:propBUps}

(i) We have
\bees
\cup_{\GV\in\CV}\Bxi(\GV) =\R^d.
\enes

(ii) $\bxi\in\Bxi_1(\GV)$ iff $\bxi_{\GV}\in\Omega(\GV)$, where $\Omega(\GV)\subset\GV$
is a certain bounded set (more precisely, $\Omega(\GV) =\Bxi_1(\GV)\cap\GV\subset \CB(m L_m)$ if
$\dim\GV =m$).

(iii) $\Bxi_1(\R^d) =\Bxi(\R^d)$ is a bounded set, $\Bxi(\R^d)\subset \CB(d L_d)$; all other sets
$\Bxi_1(\GV)$ are unbounded.
\enl

Now we move to slightly less obvious properties. From now on we always assume that $\rho_0$
(and thus $\rho_n$) is sufficiently large. We also assume, as we always do, that the value of $k$
is sufficiently large so that, for example, $L_j\rho_n^{0+}< L_{j+ 1}$.

\bel\label{lem:intersect}
Let $\GV, \GU\in\CV$. Then $\big(\Bxi_1(\GV)\cap\Bxi_1(\GU)\big)\subset \Bxi_1(\GW)$, where $\GW:= \GV+ \GU$ (algebraic sum).
\enl

\bec
(i) We can re-write definition \eqref{eq:Bxi} like this:
\bees%\label{eq:Bxibis}
\Bxi(\GV) :=\Bxi_1(\GV)\setminus\big(\cup_{\GU\not\subset\GV}\Bxi_1(\GU)\big).
\enes

(ii) If $\GV\ne\GU$, then $\Bxi(\GV)\cap\Bxi(\GU) =\emptyset$.

(iii) We have $\R^d =\sqcup_{\GV\in\CV}\Bxi(\GV)$ (the disjoint union).
\enc

\bel\label{lem:verynew}
Let $\GV\in\CV_m$ and $\GV\subset\GW\in\CV_{m +1}$. Let $\bmu$ be (any) unit vector from $\GW\ominus\GV$.
Then, for $\bxi\in\Bxi_1(\GV)$, we have  $\bxi\in\Bxi_1(\GW)$ if and only if the estimate
$\big|\lu\bxi,\bmu\ru\big|= \big|\lu\bxi_{\GV^{\perp}},\bmu\ru\big|\le L_{m +1}$ holds.
\enl

\bel\label{lem:newXi}
We have
\bees%\label{eq:newBxi}
\Bxi_1(\GV)\cap\cup_{\GU\supsetneq\GV}\Bxi_1(\GU)= \Bxi_1(\GV)\cap\cup_{\GW\supsetneq\GV, \ \dim\GW= 1+ \dim\GV}\Bxi_1(\GW).
\enes
\enl

\bec
We can re-write \eqref{eq:Bxi} as
\bee\label{eq:Bxibbis}
\Bxi(\GV):=\Bxi_1(\GV)\setminus\big(\cup_{\GW\supsetneq\GV, \dim\GW =1 +\dim\GV}\Bxi_1(\GW)\big).
\ene
\enc

\bel\label{lem:Upsilon}
Let $\GV\in\CV$ and $\bth\in\Bth_{\tilde k}$. Suppose that $\bxi\in\Bxi(\GV)$ and both points $\bxi$ and $\bxi+\bth$ are inside $\Lambda(\bth)$. Then $\bth\in\GV$ and $\bxi+\bth\in\Bxi(\GV)$.
\enl

\begin{defn}
\label{reachability:defn}
Let $\bth, \bth_1, \bth_2, \dots, \bth_l$ be some vectors from $\Bth'_{\tilde k}$,
which are not necessarily distinct.

\begin{enumerate}
\item \label{1}
We say that two vectors
$\bxi, \boldeta\in\R^d$ are \textsl{$\bth$-resonant congruent} if both $\bxi$ and $\boldeta$ are inside $\L(\bth)$ and $(\bxi - \boldeta) =l\bth$ with $l\in\Z$.
In this case we write $\bxi \leftrightarrow \boldeta \mod \bth$.
\item\label{2}
For each $\bxi\in\R^d$ we denote by $\BUps_{\bth}(\bxi)$ the set of all points which are $\bth$-resonant congruent to $\bxi$.
For $\bth\not = \mathbf 0$ we say that $\BUps_{\bth}(\bxi) = \varnothing$ if $\bxi\notin\L(\bth)$.
\item\label{3}
We say that $\bxi$ and $\boldeta$ are \textsl{$\bth_1, \bth_2, \dots, \bth_l$-resonant congruent}, if there exists a sequence $\bxi_j\in\R^d, j=0, 1, \dots, l$ such that $\bxi_0 = \bxi$, $\bxi_l = \boldeta$, and $\bxi_j \in\BUps_{\bth_j}(\bxi_{j-1})$ for $j =1, 2, \dots, l$.
\item
We say that $\boldeta\in\R^d$ and $\bxi\in\R^d$ are \textsl{resonant congruent}, if either $\bxi=\boldeta$ or $\bxi$ and $\boldeta$ are $\bth_1, \bth_2, \dots, \bth_l$-resonant congruent with some $\bth_1, \bth_2, \dots, \bth_l \in\Bth_{\tilde k}'$. The set of \textbf{all} points, resonant congruent to $\bxi$, is denoted by $\BUps(\bxi)$.
For points $\boldeta\in\BUps(\bxi)$ (note that this condition is equivalent to $\bxi\in\BUps(\boldeta)$) we write $\boldeta\leftrightarrow\bxi$.
\end{enumerate}
\end{defn}

Note that $\BUps(\bxi)= \{\bxi\}$ for any $\bxi\in\Bxi(\GX)$. Now Lemma \ref{lem:Upsilon} immediately implies
\bec\label{cor:Upsilon}
For each $\bxi\in\Bxi(\GV)$ we have $\BUps(\bxi)\subset\Bxi(\GV)$ and thus
\bees
\Bxi(\GV)= \sqcup_{\bxi\in\Bxi(\GV)}\BUps(\bxi).
\enes
\enc
\bel\label{lem:diameter}
The diameter of $\BUps(\bxi)$ is bounded above by $mL_m$, if $\bxi\in\Bxi(\GV)$, $\GV\in\CV_m$.
\enl

\bel\label{lem:finiteBUps}
For each $\bxi\in\Bxi(\GV),\ \GV\ne\R^d$, the set $\BUps(\bxi)$ is finite, and $\card\BUps(\bxi)$ is bounded uniformly in $\bxi\in\Rd\setminus\BXi(\Rd)$.
\enl

\section{Description of the approach}\label{description section}

We first prove \eqref{eq:main_lem1} assuming that the symbol $b$ of $B$ is replaced by $\tilde b$ which satisfies \eqref{tilde b}. In particular, it belongs to the class $\BS_\al$. At the end, in Section~\ref{final section}, we will use \eqref{eq:condB2} to show that Theorem~\ref{main_thm} holds as stated.

For any set $\CC\subset\R^d$ by $\CP(\CC)$ we denote  the orthogonal projection onto $\mathrm{span}\{\be_{\bxi}\}_{\bxi\in\CC}$ in $\plainB_2(\R^d)$ and by $\CP^{L}(\CC)$ the same projection considered in $\plainL2(\R^d)$, i.e.
\bee\label{CP}
\CP^{L}(\CC)=\CF^*\Id_{\CC}\CF,
\ene
where $\CF$ is the Fourier transform and $\Id_{\CC}$ is the operator of multiplication by the indicator function of $\CC$.
Obviously, $\CP^{L}(\CC)$ is a well-defined (respectively, non-zero) projection iff $\CC$ is measurable (respectively, has non-zero measure).
Let us fix sufficiently large $n$, and denote (recall that $\la_n =\rho_n^{2w}$)
\bee\label{X}
\CX_n:= \Big\{\bxi\in\R^d,\,|\bxi|^{2w}\in \big[(5/6)^{2w}\la_n, 5^{2w}\la_n\big]\Big\}.
\ene
We also put
\bees
\CA= \CA_n:= \cup_{\bxi\in\CX_n}\BUps(\bxi).
\enes
Lemma \ref{lem:diameter} implies that, if $\rho_0$ is big enough,
\bee\label{bxi in CA}
\textrm{for each $\bxi\in\CA$ we have $|\bxi|^{2w}\in\big[(2/3)^{2w}\la_n, 6^{2w}\la_n\big]$.}
\ene
In particular, we have
\bee\label{eq:CARd}
\CA\cap\BXi(\R^d)= \varnothing.
\ene
Let us define
\bees
\hat\CA :=\big\{\bxi\not\in\CA,\ |\bxi|^{2w} <\la_n\big\}
\enes
and
\bee\label{check A}
\check\CA:=\big\{\bxi\not\in\CA,\ |\bxi|^{2w}>\la_n\big\}.
\ene

We now plan to apply the gauge transform as in Sections 8 and 9 of \cite{ParSht2} to the operator $H$.
The details of this procedure will be explained in Sections~\ref{Partition section} and~\ref{Gauge transform section};
here, we just mention that we are going to introduce two operators: $H_1$ and $H_2$. The operator $H_1$ is unitary
equivalent to $H$: $H_1= U^{-1}HU$, where $U=e^{i\Psi}$ with a bounded pseudo-differential operator $\Psi$ with
almost-periodic coefficients (then Lemma \ref{norms lemma} implies that the densities of states of $H$ and $H_1$ are the same). Moreover, $H_1= H_2+ R_{\tilde k}$, where
\begin{equation}\label{bound on remainder}
\|R_{\tilde k}\|\lesssim \rho_n^{-M+ 2w- d}
\end{equation}
and $H_2= (-\Delta)^{w}+ W_{\tilde k}$ is a self-adjoint
pseudo-differential operator with symbol $|\bxi|^{2w}+ w_{\tilde
k}(\bx, \bxi)$ which satisfies the following property:
\bee\label{eq:b3} \hat w_{\tilde k}(\bth, \bxi)= 0, \ \mathrm{if} \
\big(\bxi\not\in\La(\bth)\ \&\ \bxi\in\CA\big), \  \mathrm{or} \
\big(\bxi+\bth\not\in\La(\bth)\ \&\ \bxi\in\CA\big),  \  \mathrm{or}
\ (\bth\not\in\Bth_{\tilde k}). \ene

We can now use a simple statement which follows from Lemma~\ref{norms lemma} and Remark~\ref{spurious remark}:

\bel\label{H1H2 lemma} Suppose, $H_1$ and $H_2$
are two  elliptic self-adjoint pseudo-differential operators with
almost-periodic coefficients such that $\|H_1- H_2\|\lesssim \rho_n^{-M+
2w- d}$. Suppose that $N(H_2; \rho^{2w})$ satisfies asymptotic
expansion \eqref{eq:main_lem1}. Then $N(H_1; \rho^{2w})$ also
satisfies \eqref{eq:main_lem1} with the same coefficients.
\enl

This means that it is enough to establish
the asymptotic expansion \eqref{eq:main_lem1} for the operator $H_2$
instead of $H$. Condition \eqref{eq:b3} implies that for each
$\bxi\in\CA$ the subspace $\CP\big(\BUps(\bxi)\big)\plainB_2(\R^d)$
is an invariant subspace of $H_2$; its dimension is finite by
Lemma~\ref{lem:finiteBUps}. We put
\bees%\label{eq:h3bxi}
H_2(\bxi):= H_2|_{\CP(\BUps(\bxi))\plainB_2(\R^d)}.
\enes
Note that the subspaces $\CP(\hat\CA)\plainB_2(\R^d)$ and $\CP(\check\CA)\plainB_2(\R^d)$ are invariant as well; by $H_2(\hat\CA)$ and $H_2(\check\CA)$ we denote the restrictions of $H_2$ to these subspaces; we also denote by $H_2(\CA)$ the restriction of $H_2$ to $\CP(\CA)\plainB_2(\R^d)$. If we consider the operator $H_2$ acting in $\plainL2(\R^d)$, then $\CP^L(\hat\CA)\plainL2(\R^d)$, $\CP^L(\check\CA)\plainL2(\R^d)$, and $\CP^L(\CA)\plainL2(\R^d)$ are still invariant subspaces.
It follows from \eqref{X} -- \eqref{check A} that $UH_2(\hat\CA)U^*< (5/6)^{2w}\la_n I$ and $UH_2(\check\CA)U^*> 5^{2w}\la_n I$.

For each $\bxi\in\CA$ the operator $H_2(\bxi)$ is a finite-dimensional self-adjoint operator, so its spectrum is purely discrete; we denote its eigenvalues (counting multiplicities) by $\la_1(\bxi)\le \la_2(\bxi)\le\dots\le \la_{\card\BUps(\bxi)}(\bxi)$. Next, we list all points $\boldeta\in\BUps(\bxi)$ in increasing order of their absolute values; thus, we have put into correspondence to each point $\boldeta\in\BUps(\bxi)$ a natural number $t= t(\boldeta)$ so that $t(\boldeta)< t(\boldeta')$ if $|\boldeta|< |\boldeta'|$. If two points $\boldeta= (\eta_1,\dots,\eta_d)$ and $\boldeta'= (\eta'_1, \dots, \eta'_d)$ have the same absolute values, we put them in the lexicographic order of their coordinates, i.e. we say that $t(\boldeta)< t(\boldeta')$ if $\eta_1< \eta'_1$, or $\eta_1= \eta'_1$ and $\eta_2< \eta'_2$, etc. Now we define the map $g: \CA\to \R$ which to each point $\boldeta\in\CA$ brings into correspondence the number $\la_{t(\boldeta)}\big(\BUps(\boldeta)\big)$. This map is an injection from $\CA$ onto the set of eigenvalues of $H_2$, counting multiplicities (recall that we consider the operator $H_2$ acting in $\plainB_2(\R^d)$, so there is nothing miraculous about its spectrum consisting of eigenvalues and their limit points). Moreover, all eigenvalues of $H_2$ inside the interval $\big[(7/8)^{2w}\la_n, (9/2)^{2w}\la_n\big]$ have a pre-image under $g$.
We define
\begin{equation}\label{g outside A}
g(\bxi):= |\bxi|^{2w}, \quad\textrm{for} \quad \bxi\in\R^d\setminus \CA.
\end{equation}
Arguments similar to the ones used in \cite{ParSob} show that $g$ is a measurable function.

We introduce
\bees
%\label{eq:mla}
\CG_{\la}:= \big\{\bxi\in\R^d,\,g(\bxi)\le\la\big\}.
\enes

\bel\label{volume lemma}
For $\la\in[\la_n, 4^{2w}\la_n]$ being a continuity point of $N(\la; H_2)$
we have:
\bee\label{eq:densityh3}
N(\la; H_2)= (2\pi)^{-d}\vol \CG_{\la}.
\ene
\enl

Since points of continuity of $N(\la)$ are dense, {\it the asymptotic expansion proven for such $\lambda$ can be extended to all $\la\in[\la_n, 4^{2w}\la_n]$ by taking the limit}. Thus, our next task is to compute $\vol \CG_{\la}$.
Let us put
\bees%\label{eq:47}
\CA^+(\rho):= \big\{\bxi\in\R^d,\,g(\bxi)<\rho^{2w}<|\bxi|^{2w}\big\}
\enes
and
\bees%\label{eq:48}
\CA^-(\rho):= \big\{\bxi\in\R^d,\,|\bxi|^{2w}<\rho^{2w}<g(\bxi)\big\}.
\enes

\bel\label{lem:new1}
\bee\label{eq:46}
\vol(\CG_{\la})= \om_d\rho^d+ \vol\CA^+(\rho)- \vol\CA^-(\rho),
\ene
where $\omega_d$ is the volume of the unit ball in $\R^d$.
\enl
\bep
We obviously have $\CG_{\la}= \CB(\rho)\cup \CA^+(\rho)\setminus \CA^-(\rho)$. Since $\CA^-(\rho)\subset \CB(\rho)$
and $\CA^+(\rho)\cap \CB(\rho)=\emptyset$, this implies \eqref{eq:46}.
\enp
\ber\label{rem:nnew1}
Properties of the mapping $g$ imply that $\CA^+(\rho)\cup\CA^-(\rho)\subset \CA$. Thus, in order to compute $N(\lambda)$, we need to analyze the behavior of $g$ only inside $\CA$.
\enr
We will compute volumes of $\CA^{\pm}(\rho)$ by means of integrating their characteristic functions in a specially chosen set of
coordinates. The next section is devoted to introducing these coordinates.

\section{Coordinates}\label{coordinates section}

In this section, we do some preparatory work before computing $\vol\CA^{\pm}(\rho)$. Namely, we are going to introduce a convenient set of coordinates in $\Bxi(\GV)$.
Let $\GV\in\CV_m$ be fixed; since $\CA^{\pm}(\rho)\cap\BXi(\R^d)= \emptyset$, we will assume that $m<d$. Then, as we have seen, $\bxi\in\Bxi_1(\GV)$ if and only if $\bxi_{\GV}\in\Om(\GV)$. Let $\{\GU_j\}$ be a collection of all subspaces $\GU_j\in\CV_{m+ 1}$ such that each $\GU_j$ contains $\GV$. Let $\bmu_j= \bmu_j(\GV)$ be (any) unit vector from $\GU_j\ominus\GV$.
Then it follows from Lemma~\ref{lem:verynew} that for $\bxi\in\Bxi_1(\GV)$, we have  $\bxi\in\Bxi_1(\GU_j)$ if and only if the estimate $\big|\lu\bxi, \bmu_j\ru\big|= \big|\lu\bxi_{\GV^{\perp}}, \bmu_j\ru\big|\le L_{m+ 1}$ holds. Thus, formula \eqref{eq:Bxibbis} implies that
\bees%\label{eq:Bxibbbis}
\Bxi(\GV)= \Big\{\bxi\in\R^d,\ \bxi_{\GV}\in\Omega(\GV)\ \&\ \forall j \ \big|\lu\bxi_{\GV^{\perp}},\bmu_j(\GV)\ru\big| > L_{m+ 1}\Big\}.
\enes
The collection $\big\{\bmu_j(\GV)\big\}$ obviously coincides with
\bees
\big\{\bn(\bth_{\GV^{\perp}}),\ \bth\in\Bth_{\tilde k}\setminus\GV\big\}.
\enes

The set $\Bxi(\GV)$ is, in general, disconnected; it consists of several connected components which we will denote by $\big\{\Bxi(\GV)_p\big\}_{p=1}^P$. Let us fix a connected component $\Bxi(\GV)_p$.
Then for some vectors $\big\{\tilde\bmu_j(p)\big\}_{j=1}^{J_p}\subset \{\pm\bmu_j\}$ we have
\bees%\label{eq:Bxip}
\Bxi(\GV)_p= \big\{\bxi\in\R^d,\ \bxi_{\GV}\in\Omega(\GV)\ \&\ \forall j \ \lu\bxi_{\GV^{\perp}},\tilde\bmu_j(p)\ru > L_{m +1}\big\};
\enes
we assume that $\big\{\tilde\bmu_j(p)\big\}_{j =1}^{J_p}$ is the minimal set with this property, so that each hyperplane
$$
\big\{\bxi\in\R^d,\ \bxi_{\GV}\in\Omega(\GV)\ \ \&\ \ \lu\bxi_{\GV^{\perp}}, \tilde\bmu_j(p)\ru= L_{m+ 1}\big\},\ j= 1, \dots, J_p
$$
has a non-empty intersection with the boundary of $\Bxi(\GV)_p$. It is not hard to see that $J_p\ge d- m$. Indeed, otherwise $\Bxi(\GV)_p$ would have non-empty intersection with $\Bxi_1(\GV')$ for some $\GV'$, $\GV\subsetneq\GV'$. We also introduce
\bees%\label{eq:Bxiptilde}
\tilde\Bxi(\GV)_p:= \big\{\bxi\in\GV^{\perp},\ \forall j \ \lu\bxi,\tilde\bmu_j(p)\ru > 0\big\}.
\enes
Note that our assumption that $\Bxi(\GV)_p$ is a connected component of $\Bxi(\GV)$ implies that for any $\bxi\in\tilde\Bxi(\GV)_p$ and any $\bth\in\Bth_{\tilde k}\setminus\GV$ we have
\bees%\label{eq:ne0}
\lu\bxi,\bth\ru= \lu\bxi,\bth_{\GV^{\perp}}\ru\ne 0.
\enes
We also put
\begin{equation*}
K:= d- m- 1.
\end{equation*}

Without loss of generality we may (and will) assume that the number $J_p$ of `defining planes' is the minimal possible, i.e. $J_p= K+ 1$. Indeed, the argument presented in Section~11 of \cite{ParSht2} explains
how to derive the result for arbitrary $\BXi(\GV)_p$, assuming we have
proved it in the case $J_p= K+ 1$.

If $J_p= K+ 1$, then the set $\big\{\tilde\bmu_j(p)\big\}_{j= 1}^{K+ 1}$ is linearly independent.
Let $\ba= \ba(p)$ be a unique point from $\GV^\perp$ satisfying the following conditions: $\lu\ba,\tilde\bmu_j(p)\ru= L_{m+ 1}$, $j= 1, \dots, K+ 1$.
Then, since the determinant of the Gram matrix of vectors $\tilde\bmu_j(p)$ is $\gtrsim\rho_n^{0-}$ by \eqref{eq:condC1}, we have
\bee\label{bound on a}
|\ba|\lesssim L_{m+ 1}\rho_n^{0+}= \rho_n^{\al_{m+ 1}+ 0+}.
\ene
We introduce shifted cylindrical coordinates in $\Bxi(\GV)_p$. These coordinates will be denoted by $\bxi= (r; \mathbf\Phi; \mathbf X)$. Here, $\mathbf X= (X_1, \dots, X_m)$ is an arbitrary set of cartesian coordinates in $\Omega(\GV)$. These coordinates do not depend on the choice of the connected component $\Bxi(\GV)_p$. The rest of the coordinates $(r, \mathbf\Phi)$ are shifted spherical coordinates in $\GV^{\perp}$, centered at $\ba$. This means that
\bees%\label{eq:r}
r(\bxi)= |\bxi_{\GV^{\perp}}- \ba|
\enes
and
\bees%\label{eq:tildePhi}
\mathbf\Phi= \bn(\bxi_{\GV^{\perp}}- \ba)\in S_{\GV^{\perp}}.
\enes
More precisely, $\mathbf\Phi\in \CM_p$, where $\CM_p:= \big\{\bn(\bxi_{\GV^{\perp}}- \ba),\ \bxi\in\Bxi(\GV)_p\big\}\subset S_{\GV^{\perp}}$ is a $K$-dimensional spherical simplex with $K+ 1$ sides. Note that
\bees%\label{eq:Mp}
\bes
\CM_p&= \big\{\bn(\bxi_{\GV^{\perp}}-\ba),\ \bxi\in\Bxi(\GV)_p\big\}= \big\{\bn(\bxi_{\GV^{\perp}}-\ba),\  \forall j \ \lu\bxi_{\GV^{\perp}},\tilde\bmu_j(p)\ru > L_{m+1}\big\}\\
&= \big\{\bn(\boldeta),\ \boldeta:= \bxi_{\GV^{\perp}}-\ba\in\GV^\perp,\  \forall j \ \lu\boldeta,\tilde\bmu_j(p)\ru > 0\big\}= S_{\GV^{\perp}}\cap\tilde\Bxi(\GV)_p.
\end{split}
\enes
We will denote by $d\mathbf\Phi$ the spherical Lebesgue measure on $\CM_p$.
For each non-zero vector $\bmu\in\GV^{\perp}$, we denote
\bees
\CW(\bmu):= \big\{\boldeta\in\GV^\perp,\ \lu\boldeta,\bmu\ru= 0\big\}.
\enes
Thus, the sides of the simplex $\CM_p$ are intersections of $\CW\big(\tilde\bmu_j(p)\big)$ with the sphere $S_{\GV^{\perp}}$.
Each vertex $\bv= \bv_t$, $t= 1, \dots, K+ 1$ of $\CM_p$ is an intersection of $S_{\GV^{\perp}}$ with $K$ hyperplanes
$\CW\big(\tilde\bmu_j(p)\big)$, $j= 1, \dots, K+ 1$, $j\ne t$. This means that $\bv_t$ is a unit vector from $\GV^{\perp}$ which is orthogonal to $\big\{\tilde\bmu_j(p)\big\}$, $j= 1, \dots, K+ 1$, $j\ne t$; this defines $\bv$ up to a multiplication by $-1$.

\bel\label{lem:newangles}
Let $\GU_1$ and $\GU_2$ be two strongly distinct subspaces each of which is a linear combination of some of the
vectors from $\big\{\tilde\bmu_j(p)\big\}$. Then the angle between them is not smaller than $s(\rho_n)$.
In particular, all non-zero angles between two sides of any dimensions of $\CM_p$ as well as all the distances between two vertexes $\bv_t$ and $\bv_{\tau}$, $t\ne\tau$, are bounded below by $s(\rho_n)$.
\enl

\bel\label{lem:sign}
Let $p$ be fixed.
Suppose, $\bth\in\Bth_{\tilde k}\setminus\GV$ and $\bth_{\GV^{\perp}}=\sum_{j=1}^{K+1} b_j\tilde\bmu_j(p)$.
Then either all coefficients $b_j$ are non-positive, or all of them are non-negative.
\enl

By taking sufficiently large $\tilde k$ we can assure that the diameter of $\CM_p$ does not exceed $(100d^2)^{-1}$. We put $\Phi_q:= \frac{\pi}{2}-\phi\big(\bxi_{\GV^\perp}- \ba, \tilde\bmu_q(p)\big)$, $q= 1, \dots, K+ 1$.
The geometrical meaning of these coordinates is simple: $\Phi_q$ is the spherical distance between
$\mathbf\Phi= \bn(\bxi_{\GV^{\perp}}-\ba)$ and $\CW\big(\tilde\bmu_q(p)\big)$. The reason why we have introduced $\Phi_q$ is that in these coordinates some important objects will be especially simple (see e.g. Lemma~\ref{lem:products} below) which is very convenient for integration. At the same time, the set of coordinates $\big(r, \{\Phi_q\}\big)$ contains $K+ 2$ variables, whereas we only need $K+ 1$ coordinates in $\GV^{\perp}$. Thus, we have one constraint for variables $\Phi_j$. Namely, let $\{\be_j\}$, $j= 1, \dots, K+ 1$ be a fixed orthonormal basis in $\GV^{\perp}$ chosen in such a way that the $K+1$-st axis is directed along $\ba$, and thus passes through $\CM_p$.
Then we have $\be_j= \sum_{l= 1}^{K+ 1}a_{jl}\tilde\bmu_l$ with some matrix $\{a_{jl}\}$, $j,l= 1, \dots, K+ 1$, and $\tilde\bmu_l= \tilde\bmu_l(p)$.
Therefore (recall that we denote $\boldeta:= \bxi_{\GV^{\perp}}- \ba$),
\bees%\label{eq:etaj}
\eta_j= \lu\boldeta,\be_j\ru= r\sum_{q= 1}^{K+ 1}a_{jq}\sin\Phi_q
\enes
and, since $r^2(\bxi)=|\boldeta|^2=\sum_{j=1}^{K+1}\eta_j^2$, this
implies that
\bees%\label{eq:odin}
\sum_{j= 1}^{K+ 1}\Big(\sum_{q= 1}^{K+ 1} a_{jq}\sin\Phi_q\Big)^2= 1,
\enes
which is our constraint.

Let us also put
\bee\label{eq:etajdash}
\eta_j':= \frac{\eta_j}{|\boldeta|}= \sum_{q= 1}^{K+ 1}a_{jq}\sin\Phi_q.
\ene
Then we can write the surface element $d\mathbf\Phi$ in the coordinates $\{\eta_j'\}$ as
\bees%\label{eq:surfaceelement}
d\mathbf\Phi= \frac{d\eta_1'\dots d\eta_K'}{\eta_{K+ 1}}= \frac{d\eta_1'\dots d\eta_K'}{\big(1- \sum_{j= 1}^K(\eta_{j}')^2\big)^{1/2}},
\enes
where the denominator is bounded below by $1/2$ by our choice of the basis $\{\be_j\}$.
It follows from our choice of the coordinates and \eqref{eq:etajdash} that
\begin{equation}\label{a_dot_eta}
\lu\ba, \mathbf\Phi\ru= \lu\ba, \bn(\boldeta)\ru= |\ba|\eta_{K+ 1}'= |\ba|\sum_{q= 1}^{K+ 1}a_{K+ 1\, q}\sin\Phi_q.
\end{equation}

\bel\label{lem:Al}
For each $p,l$ we have $|a_{pl}|\le s(\rho_n)^{-1}$.
\enl

\bel\label{lem:anglebelow}
We have $\max_j\sin\Phi_j(\boldeta)\ge s(\rho_n) d^{-3/2}$.
\enl

The next lemma describes the dependence on $r$ of all possible inner products $\lu\bxi,\bth\ru$, $\bth\in\Bth_{\tilde k}$, $\bxi\in\Bxi(\GV)_p$.
\bel\label{lem:products}
Let
$\bxi\in\Bxi(\GV)_p$, $\GV\in\CV_m$, and $\bth\in\Bth_{\tilde k}$.

(i) If $\bth\in\GV$, then $\lu\bxi,\bth\ru$ does not depend on $r$.

(ii) If $\bth\not\in\GV$ and $\bth_{\GV^{\perp}}=\sum_{q}b_q\tilde\bmu_q(p)$, then
\bees%\label{eq:innerproduct1}
\lu\bxi,\bth\ru=\lu \mathbf X,\bth_{\GV}\ru+L_{m+1}\sum_{q}b_q+r(\bxi)\sum_{q}b_q\sin\Phi_q.
\enes

In the case (ii) all the coefficients $b_q$ are either non-positive or non-negative and
each non-zero coefficient $b_q$ satisfies
\bee\label{eq:n10}
\rho_n^{0-}\lesssim |b_q| \lesssim \rho_n^{0+}.
\ene
\enl

\section{Partition of the perturbation}\label{Partition section}

The symbols we are going to construct in this section will depend on $\rho_n$; this dependence will usually be omitted from the notation.

Let $\varpi\in \plainC\infty(\R)$ be such that
\begin{equation}\label{eta:eq}
0\le\varpi\le 1,\ \ \varpi(z)=
\begin{cases}
& 1,\  z \le 1;\\
& 0,\  z \ge 21/20.
\end{cases}
\end{equation}
For $\bth\in \Bth'$ we define several $\plainC\infty$-cut-off functions:
\begin{equation}\label{el:eq}
\begin{cases}
e_{\bth}(\bxi)&:= \varpi\Big(\big|2|2\bxi+ \bth|/\rho_n- 15\big|/13\Big),\\
\ell^{>}_{\bth}(\bxi)&:= 1- \varpi\Big(\big(2|2\bxi+ \bth|/\rho_n- 15\big)/13\Big),\\
\ell^{<}_{\bth}(\bxi)&:= 1- \varpi\Big(\big(15- 2|2\bxi+ \bth|/\rho_n\big)/13\Big),
\end{cases}
\end{equation}
and
\begin{equation}\label{phizeta:eq}
\begin{cases}
\z_{\bth}(\bxi)&:= \varpi\biggl(\dfrac{\big|\lu\bth, \bxi+ \bth/2\ru\big|}{\rho_n^\beta|\bth|}\biggr),\\
\varphi_{\bth}(\bxi)&:= 1- \z_{\bth}(\bxi).
\end{cases}
\end{equation}

\ber\label{partition support remark}
Note that $e_{\bth}+ \ell^{>}_{\bth}+ \ell^{<}_{\bth}= 1$.
The function $\ell^{>}_{\bth}$ is supported on the set $|\bxi+ \bth/2|\geqslant 7\rho_n$, and $\ell^{<}_{\bth}$ is supported on the set $|\bxi+ \bth/2|\leqslant \rho_n/2$. The function $e_{\bth}$ is supported in the shell $\rho_n/3\le |\bxi+ \bth/2|\le 8\rho_n$.
\enr

Using the notation $\ell_{\bth}$ for any of the functions $\ell^{>}_{\bth}$ or $\ell^{<}_{\bth}$, we point out that
\begin{equation*}%\label{symmetry:eq}
\begin{cases}
e_{\bth}(\bxi)= e_{-\bth}(\bxi+ \bth), &\ell_{\bth}(\bxi)= \ell_{-\bth}(\bxi+ \bth),\\
\varphi_{\bth}(\bxi)=  \varphi_{-\bth}(\bxi+ \bth), &\z_{\bth}(\bxi)= \z_{-\bth}(\bxi+ \bth).
\end{cases}
\end{equation*}
Note that the above functions satisfy the estimates
\begin{equation}\label{varphi:eq}
\begin{cases}
\big|\BD^{\bs}_{\bxi}e_{\bth}(\bxi)\big|+ \big|\BD^{\bs}_{\bxi}\ell_{\bth}(\bxi)\big|\lesssim \rho_n^{-|\mathbf s|},\\
\big|\BD^{\bs}_{\bxi}\varphi_{\bth}(\bxi)\big|+ \big|\BD^{\bs}_{\bxi}\z_{\bth}(\bxi)\big|\lesssim \rho_n^{-\beta|\bs|}.
\end{cases}
\end{equation}
Now for any symbol $b\in\BS_{\a}(\b)$ we introduce five new symbols:
\begin{equation*}
\begin{split}
b^{\ssharp}(\bx, \bxi; \rho_n)&:= \sum_{\bth\in\Bth'}\hat b(\bth, \bxi)\ell^{>}_{\bth}(\bxi)e^{i\bth \bx}, \\
b^{\natural}(\bx, \bxi; \rho_n)&:= \sum_{\bth\in\Bth'}\hat b(\bth, \bxi)\varphi_{\bth}(\bxi)e_{\bth}(\bxi) e^{i\bth \bx}, \\
b^{\flat}(\bx, \bxi; \rho_n)&:= \sum_{\bth\in\Bth'}\hat b(\bth, \bxi)\z_{\bth}(\bxi)e_{\bth}(\bxi)e^{i\bth \bx}, \\
b^{\downarrow}(\bx, \bxi; \rho_n)&:= \sum_{\bth\in\Bth'}\hat b(\bth, \bxi)\ell^{<}_{\bth}(\bxi)e^{i\bth \bx}, \\
b^o(\bx, \bxi; \rho_n)&= b^o(\bxi; \rho_n):= \hat b(0, \bxi).
\end{split}
\end{equation*}
The superscripts here are chosen to mean, respectively: `large energy', `non-resonant', `resonant', `small energy' and $0$-th Fourier coefficient. The corresponding operators are denoted by $B^{\ssharp}$, $B^{\natural}$, $B^{\flat}$, $B^{\downarrow}$, and $B^o$.
By definitions \eqref{eta:eq}, \eqref{el:eq} and \eqref{phizeta:eq}
\begin{equation}\label{b as a sum}
b= b^o+ b^{\downarrow}+ b^{\flat}+ b^{\natural}+ b^{\ssharp}.
\end{equation}
The role of each of these operators is easy to explain. Note that on the support of the functions
$\hat b^{\natural}(\bth, \cdot; \rho_n)$ and $\hat b^{\flat}(\bth, \cdot; \rho_n)$ we have (using \eqref{bound on R})
\begin{equation*}%\label{ds:eq}
\rho_n/3 - O(\rho_n^{0+})\le |\bxi|\le 8\rho_n + O(\rho_n^{0+}).
\end{equation*}
On the support of $b^{\downarrow}(\bth, \cdot; \rho_n)$ we have
\begin{equation}\label{supportell<:eq}
|\bxi|\le \rho_n/2 + O(\rho_n^{0+}).
\end{equation}
On the support of $b^{\ssharp}(\bth, \ \cdot\ ; \rho_n)$ we have
\begin{equation}\label{supportell>:eq}
|\bxi|\ge 7\rho_n- O(\rho_n^{0+}).
\end{equation}
The introduced symbols play a central role in the proof of Lemma~\ref{main_lem}. As we have seen in Section~\ref{description section}, due to \eqref{supportell<:eq} and \eqref{supportell>:eq} the symbols $b^\downarrow$ and $b^\ssharp$ make only a negligible contribution to the spectrum of the operator $H$ near $\l= \rho^{2w}$ for $\rho\in I_n$. The only significant components of $b$ are the symbols $b^\natural, b^{\flat}$ and $b^o$. The symbol $b^o$ will remain as it is, and the symbol $b^\natural$ will be transformed in the next section to another symbol, independent of $\bx$.

Under the condition $b\in\BS_{\a}(\b)$ the above symbols belong to the same class $\BS_{\a}(\b)$ and the following bounds hold:
\begin{equation*}%\label{subord:eq}
\1 b^{\flat}\1^{(\a)}_{l, s}+ \1 b^{\natural}\1^{(\a)}_{l, s}+ \1 b^{\ssharp}\1^{(\a)}_{l, s}+ \1 b^{o}\1^{(\a)}_{l, s}+ \1 b^{\downarrow}\1^{(\a)}_{l, s}\lesssim \1 b\1^{(\a)}_{l, s}.
\end{equation*}
If $b$ symmetric, then so are the symbols on the right hand side of \eqref{b as a sum}.

Let us mention some other elementary properties of the introduced operators. In the lemma below we use the projection $\CP(\CC)$, $\CC\subset\Rd$ which was defined in Section~\ref{description section}.

\begin{lem}\label{smallorthog:lem}
Let $b\in \BS_{\a}(\b)$ with
some $\a\in\R$. Then:
\begin{itemize}
\item[(i)]
The operator $B^{\downarrow}$ is bounded and
\begin{equation*}
\|B^{\downarrow}\|\lesssim \1 b \1^{(\a)}_{0, 0} \rho_n^{\b\max(\a,
0)}.
\end{equation*}
Moreover,
\begin{equation*}
\Big(I- \CP\big(\CB(2\rho_n/3)\big)\Big) B^{\downarrow}= B^{\downarrow}\Big(I- \CP \big(\CB(2\rho_n/3)\big)\Big)= 0.
\end{equation*}

\item[(ii)] The operator $B^\flat$ satisfies the relations
\begin{equation*}%\label{bflatorthog:eq}
\CP\big(\CB(\rho_n/6)\big)B^{\flat}= B^{\flat}\CP\big(\CB(\rho_n/6)\big)= \Big(I- \CP\big(\CB(9\rho_n)\big)\Big)B^{\flat}= B^{\flat}\Big(I- \CP\big(9\rho_n)\big)\Big)= 0,
\end{equation*}
and similar relations hold for the operator $B^{\natural}$ as well.
Moreover, $b^{\natural}, b^{\flat}\in \BS_{\g}$ for any $\g\in \R$, and for all $l$ and $s$
\begin{equation*}%\label{nat:eq}
\1 b^{\natural}\1^{(\g)}_{l, s} + \1 b^{\flat}\1^{(\g)}_{l, s} \lesssim
\rho_n^{\b(\a - \g)}\1 b\1^{(\a)}_{l, s},
\end{equation*}
with the implied constant independent of $b$ and $n\ge 1$. In particular, the operators $B^{\natural}, B^{\flat}$ are bounded and
\begin{equation*}
\|B^{\natural}\|+ \|B^{\flat}\|\lesssim \rho_n^{\b\a}\1 b\1^{(\a)}_{0, 0}.
\end{equation*}

\item[(iii)]
\begin{equation*}
\CP\bigl(\CB(6\rho_n)\bigr)B^{\ssharp}= B^{\ssharp}\CP\bigl(\CB(6\rho_n)\bigr) = 0.
\end{equation*}
\end{itemize}
\end{lem}

\section{Operators $H_1$ and $H_2$}\label{Gauge transform section}

\subsection{Preparation}
As mentioned at the end of Section~\ref{reduction section}, we assume that the symbol $b$ of $B$ satisfies \eqref{tilde b}, and thus belongs to the class $\BS_\al(\b)$ with $\alpha$ defined in \eqref{alpha}.
Our strategy is to find a unitary operator which reduces $H= H_0+ B$, $H_0:= (-\Delta)^w$, to another PDO, whose symbol, essentially, depends only on $\bxi$. More precisely, we want to find operators $H_1$ and $H_2$ with the properties discussed in Section~\ref{description section}.

Repeating the calculations of Subsection~9.1 of \cite{ParSht2} we find that $H$ is unitarily equivalent to
\begin{equation}\label{H_1}
H_1 =H_0 +Y^{(o)}_{\tilde k} +Y_{\tilde k}^{\flat} +Y_{{\tilde k}}^{\downarrow, \ssharp} +R_{\tilde k},
\end{equation}
where
\begin{align}
Y_{\tilde k} &:=\sum_{l =1}^{{\tilde k}}B_l +\sum_{l =2}^{{\tilde k}}T_l, \label{Y_tilde k}\\
B_1 &:=\op(b),\notag\\
B_l &:=\sum_{j =1}^{l -1}\frac{1}{j!}\sum_{k_1+ k_2+ \dots+ k_j= l- 1} \ad\big(\op(b); \Psi_{k_1}, \Psi_{k_2}, \dots, \Psi_{k_j}\big),\ l\ge 2, \label{bl:eq}\\
T_l &:=\sum_{j =2}^l \frac{1}{j!} \sum_{k_1+ k_2 +\dots +k_j =l}
\ad(H_0; \Psi_{k_1}, \Psi_{k_2}, \dots, \Psi_{k_j}),\  l\ge 2
\label{tl:eq},\\
R_{\tilde k} &:=\int_0^1dt_1\int_0^{t_1}dt_2\cdots \int_0^{t_{\tilde k}}\exp(-it\Psi)\ad^{{\tilde k}+ 1}(H; \Psi)\exp(it\Psi)dt\notag \\ &+\sum_{j =1}^{\tilde k} \frac{1}{j!}\sum_{\substack{k_1+ k_2 +\dots +k_j\ge  {\tilde k} +1,\\ k_q\leqslant \tilde k, \ q =1, \dots, j}} \ad(H; \Psi_{k_1}, \Psi_{k_2}, \dots, \Psi_{k_j}),\notag\\
\Psi &:= \sum_{p =1}^{\tilde k}\Psi_p.\notag
\end{align}

The symbols $\psi_j$ of PDO $\Psi_j$ are found from the following system of commutator equations:
\begin{gather}
\ad(H_0; \Psi_1) +B_1^{\natural} =0,\label{psi1:eq}\\
\ad(H_0; \Psi_l) +B_l^{\natural} +T_l^{\natural} =0,\ l\ge 2.\label{psil:eq}
\end{gather}
By Lemma~\ref{smallorthog:lem}(ii), the operators $B_l^{\natural}$, $T_l^{\natural}$ are bounded. This, in view of \eqref{psi1:eq} and \eqref{psil:eq}, implies boudedness of the commutators $\ad(H_0; \Psi_l)$, $l\geqslant 1$.
Below we denote by $y_{\tilde k}$ the symbol of the PDO $Y_{\tilde k}$.

\subsection{Commutator equations}
Put
\begin{equation}\label{chi}
\tilde{\chi}_{\bth}(\bxi) :=e_{\bth}(\bxi)\varphi_{\bth}(\bxi)\big(|\bxi +\bth|^{2w} -|\bxi|^{2w}\big)^{-1}
\end{equation}
when $\bth\not ={\bf 0}$, and $\tilde{\chi}_{\bf 0}(\bxi) :=0$.

We have

\begin{lem} \label{commut:lem}
Let $A =\op(a)$ be a symmetric PDO with $a\in\BS_{\om}$.
Then the PDO $\Psi$ with the Fourier coefficients of the symbol $\psi(\bx, \bxi)$ given by
\begin{equation}\label{psihat:eq}
\hat\psi(\bth, \bxi) :=i\,{\hat a}(\bth, \bxi)\tilde{\chi}_{\bth}(\bxi)
\end{equation}
solves the equation
\begin{equation*}%\label{adb:eq}
\ad(H_0; \Psi) +\op(a^{\natural})= 0.
\end{equation*}
Moreover, the operator $\Psi$ is bounded and self-adjoint, its symbol $\psi$ belongs to $\BS_{ \g}$ with any $\g \in\R$ and the following bound holds:
\begin{equation*}%\label{psitau:eq}
\1\psi\1^{(\g)}_{l, s} \lesssim\rho_n^{\b(\om -\g -1)- 2w+ 2}r(\rho_n)^{-1}\1 a\1^{(\om)}_{l -1, s} \lesssim\rho_n^{\b(\om -\g -1)- 2w+ 2+ 0+}\1 a\1^{(\om)}_{l -1, s}.
\end{equation*}
\end{lem}

The proof of this lemma is analogous to that of Lemma~4.1 of \cite{ParSob} and is based on the estimate
\begin{equation*}%\label{denominator control}
|\bxi +\bth|^{2w} -|\bxi|^{2w}= |\bxi|^{2w}\Big(\big(1+ |\bxi|^{-2}(2\bxi+ \bth)\cdot\bth\big)^w- 1\Big)\asymp \rho^{2w- 2}\big|\bth\cdot(\bxi+ \bth/2)\big|
\end{equation*}
which holds for $\bxi$ in the support of $e_{\bth}\varphi_{\bth}$.

Using Propositions~\ref{bound:prop}, \ref{product:prop},
\ref{commut0:prop}, Lemma~\ref{commut:lem}, and repeating arguments
from the proof of Lemma 4.2 from \cite{ParSob} (with $\sigma_j:=
j\big(\a- 2- (2w- 2)\b^{-1}\big)+ 1$), we obtain the following
\bel\label{estimateskm:lem} Let $b\in\BS_{\al}(\b)$ be a symmetric
symbol. Suppose that $k$ is large enough so that
$r(\rho_n)^{-1}\lesssim\rho_n^{0+}\lesssim\rho_n^{w+ \b- \frac{\al\b}2 -1}$
and $\tilde k$ satisfies \bee\label{eq:kM} {\tilde k}> 2(M+ \al\b+
d- 2w)/(2w+ 2\b- \al\b -2). \ene Then $\psi_j,\,
b_j,\,t_j\in\BS_\gamma(\beta)$ for any $\gamma\in\R$ and there
exists sufficiently large $\rho_0$, such that
\begin{equation}\label{R_tilde k estimate}
\|R_{\tilde k}\|\lesssim\rho_n^{-M+ 2w- d}.
\end{equation}
\enl

\ber
Note that the expression in the denominator of \eqref{eq:kM} is positive by \eqref{beta and alphas} and \eqref{alpha}.
\enr

Now Lemmas~\ref{H1H2 lemma} and \ref{estimateskm:lem} imply that the contribution of $R_{\tilde k}$ to the integrated density of states can be neglected. More precisely, let $W_{\tilde k}$ be the operator with symbol
\bee\label{eq:newy}
w_{\tilde k}(\bx, \bxi):= y_{\tilde k}(\bx, \bxi)- y_{\tilde k}^{\natural}(\bx, \bxi),\ \ \hbox{i.e.}\ \
\hat w_{\tilde k}(\bth, \bxi)= \hat y_{\tilde k}(\bth, \bxi)\big(1- e_{\bth}(\bxi)\varphi_{\bth}(\bxi)\big).
\ene
We introduce $H_2:= (-\Delta)^w+ W_{\tilde k}$. Then, by \eqref{H_1} and \eqref{R_tilde k estimate}, $\|H_1- H_2\|\lesssim\rho_n^{-M+ 2w- d}$ and, moreover, the symbol $w_{\tilde k}$ satisfies \eqref{eq:b3}. This means that all the constructions of Section~\ref{description section} are valid, and all we need to do is to compute $\vol \CG_{\lambda}$.

Until this point, the material in our paper was quite similar to the corresponding parts of \cite{ParSht2}. From now on, the differences will be substantial.

\subsection{Computing the symbol of the operator after gauge transform}
The following lemma provides us with a more explicit form of the symbol $\hat{y}_{\tilde k}$.

\bel\label{symbol} We have $\hat{y}_{\tilde k}(\bth,\bxi)=0$ for $\bth\not\in\Bth_{\tilde k}$. Otherwise,
\begin{equation}\label{symbol equation}
\begin{split}
&\hat{y}_{\tilde k}(\bth, \bxi)= \hat{b}(\bth, \bxi)\\ &+ \sum\limits_{s= 1}^{\tilde k- 1}\sum_{\substack{\bth_j, \bth_{s+ 1}\in\Bth\\ \bphi_j, \bphi_{s+ 1}, \bphi_j'\in \Bth_{s+ 1}\\ \bth_j'\in \Bth_{s+ 1}'\\ 1\leqslant j\leqslant s}}\sum_{p= 1}^s\sum_{\substack{\bth_q'', \bphi_q''\in \Bth'_{s+ 1}\\ 1\leqslant q\leqslant p- 1}} \sum_{\substack{\nu_1, \dots, \nu_{2s+ p}\geqslant 0\\ \sum\nu_i= s}}\prod_{q= 1}^{p- 1}\widehat{(\nabla^{\nu_q}e_{\bth_q''}\varphi_{\bth_q''})}(\bxi+ \bphi_q'')\\ &\times\widehat{(\nabla^{\nu_{p}}b)}(\bth_{s+ 1}, \bxi+ \bphi_{s+ 1})\prod\limits_{j= 1}^s \widehat{(\nabla^{\nu_{p+ j}}b)}(\bth_j, \bxi+ \bphi_j)\widehat{(\nabla^{\nu_{p+ s+ j}}\tilde{\chi}_{\bth_j'})}(\bxi+ \bphi_j').
\end{split}
\end{equation}
Here for $\nu\in \N$
\begin{equation}\label{Delta-power}
\nabla^\nu:= \sum_{\boldeta_1, \dots, \boldeta_\nu\in\Bth}C^{(s, p)}_{\boldeta_1, \dots, \boldeta_\nu}\big(\{\bth, \bphi\}\big)\nabla_{\boldeta_1}\cdots\nabla_{\boldeta_\nu}; \quad \nabla^0:= C^{(s, p)}\big(\{\bth, \bphi\}\big),
\end{equation}
and, for $\bth\in \R^d$, the action of $\nabla_{\bth}$ on symbols of PDO is defined in \eqref{Delta}, whereas for any function $f$ on $\Rd$
\[
(\nabla_{\bth}f)(\bxi):= f(\bxi+ \bth)- f(\bxi).
\]

The coefficients $C^{(s, p)}\big(\{\bth, \bphi\}\big)$ and $C^{(s, p)}_{\boldeta_1, \dots, \boldeta_\nu}\big(\{\bth, \bphi\}\big)$ depend on $s,\ p$ and all vectors $\bth$, $\bth_j$, $\bth_{s+ 1}$, $\bphi_j$, $\bphi_{s+ 1}$, $\bth_j'$, $\bphi_j'$, $\bth_q''$, $\bphi_q''$ (and on $\boldeta_1, \dots, \boldeta_\nu$ if these subscripts are present).
Moreover, these coefficients can differ for each particular $\nabla^\nu$, $\nu \in\N_0$.
At the same time, they are uniformly bounded by a constant which depends on $\tilde k$ only.
We apply the convention that $\prod_{q= 1}^{0}\widehat{(\nabla^{\nu_q}e_{\bth_q''}\varphi_{\bth_q''})}(\bxi+ \bphi_q'')= 1$.
\enl

\begin{proof}
We will prove the lemma by induction. Namely, let $\ell\geqslant2$. We claim that:

1) For any $m= 1, \dots, \ell- 1$, $\hat{\psi}_m(\bth, \bxi)= 0$ for $\bth\not\in\Bth_m$. Otherwise,
\begin{equation}\label{inds1}
\begin{split}
\hat{\psi}_m(\bth, \bxi)&= \sum_{\substack{\bth_j\in\Bth\\ \bphi_j, \bphi_j'\in \Bth_{m}\\ \bth_j'\in \Bth_{m}'\\ 1\leqslant j\leqslant m}}\sum_{p= 1}^{m}\sum_{\substack{\bth_q'', \bphi_q''\in \Bth'_{m}\\ 1\leqslant q\leqslant p- 1}} \sum_{\substack{\nu_1, \dots, \nu_{2m+ p- 1}\geqslant 0\\ \sum\nu_i= m- 1}}\prod_{q= 1}^{p- 1}\widehat{(\nabla^{\nu_q}e_{\bth_q''}\varphi_{\bth_q''})}(\bxi+ \bphi_q'')\\ &\times\prod\limits_{j= 1}^m \widehat{(\nabla^{\nu_{p- 1+ j}}b)}(\bth_j, \bxi+ \bphi_j)\widehat{(\nabla^{\nu_{p- 1+ m+ j}}\tilde{\chi}_{\bth_j'})}(\bxi+ \bphi_j').
\end{split}
\end{equation}

2) For any $s= 1, \dots, \ell- 1$ and any $k_1, \dots, k_p$ $(p\geqslant1)$ such that $k_1+ \dots+ k_p= s$, $\widehat{\ad\big(\op(b); \Psi_{k_1}, \dots, \Psi_{k_p}\big)}(\bth, \bxi)= 0$ for $\bth\not\in\Bth_{s+ 1}$. Otherwise,
\begin{equation}\label{inds2}
\begin{split}
&\widehat{\ad\big(\op(b); \Psi_{k_1}, \dots, \Psi_{k_p}\big)}(\bth, \bxi)\\ &= \sum_{\substack{\bth_j, \bth_{s+ 1}\in\Bth\\ \bphi_j, \bphi_{s+ 1}, \bphi_j'\in \Bth_{s+ 1}\\ \bth_j'\in \Bth_{s+ 1}'\\ 1\leqslant j\leqslant s}}\sum_{p= 1}^s\sum_{\substack{\bth_q'', \bphi_q''\in \Bth'_{s+ 1}\\ 1\leqslant q\leqslant p- 1}} \sum_{\substack{\nu_1, \dots, \nu_{2s+ p}\geqslant 0\\ \sum\nu_i= s}}\prod_{q= 1}^{p- 1}\widehat{(\nabla^{\nu_q}e_{\bth_q''}\varphi_{\bth_q''})}(\bxi+ \bphi_q'')\\ &\times\widehat{(\nabla^{\nu_{p}}b)}(\bth_{s+ 1}, \bxi+ \bphi_{s+ 1})\prod\limits_{j= 1}^s \widehat{(\nabla^{\nu_{p+ j}}b)}(\bth_j, \bxi+ \bphi_j)\widehat{(\nabla^{\nu_{p+ s+ j}}\tilde{\chi}_{\bth_j'})}(\bxi+ \bphi_j').
\end{split}
\end{equation}

3) For any $s= 2, \dots, \ell$ and any $k_1, \dots, k_p$ $(p\geqslant2)$ such that $k_1+ \dots+ k_p= s$,
$\widehat{\ad(H_0; \Psi_{k_1}, \dots, \Psi_{k_p})}(\bth, \bxi)= 0$ for $\bth\not\in\Bth_{s}$. Otherwise,
\begin{equation}\label{inds3}
\begin{split}
&\widehat{\ad(H_0; \Psi_{k_1}, \dots, \Psi_{k_p})}(\bth, \bxi)\\ &= \sum_{\substack{\bth_j, \bth_s\in\Bth\\ \bphi_j, \bphi_s, \bphi_j'\in \Bth_{s}\\ \bth_j'\in \Bth_{s}'\\ 1\leqslant j\leqslant s- 1}}\sum_{p= 1}^s\sum_{\substack{\bth_q'', \bphi_q''\in \Bth_{s}\\ 1\leqslant q\leqslant p- 1}} \sum_{\substack{\nu_1, \dots, \nu_{2s+ p- 2}\geqslant 0\\ \sum\nu_i= s- 1}}\prod_{q= 1}^{p- 1}\widehat{(\nabla^{\nu_q}e_{\bth_q''}\varphi_{\bth_q''})}(\bxi+ \bphi_q'')\\ &\times\widehat{(\nabla^{\nu_{p}}b)}(\bth_{s}, \bxi+ \bphi_{s})\prod\limits_{j= 1}^{s- 1} \widehat{(\nabla^{\nu_{p+ j}}b)}(\bth_j, \bxi+ \bphi_j)\widehat{(\nabla^{\nu_{p+ s- 1+ j}}\tilde{\chi}_{\bth_j'})}(\bxi+ \bphi_j').
\end{split}
\end{equation}

For $\ell =2$ assumptions 1)--3) can be easily checked. Indeed, by \eqref{psihat:eq}, \eqref{comm:eq} and \eqref{nabla of product},
\begin{equation*}%\label{ind1}
\hat{\psi}_1(\bth,\bxi) =i\hat{b}(\bth,\bxi)\tilde{\chi}_{\bth}(\bxi),
\end{equation*}
\begin{equation*}%\label{ind3}
\begin{split}
&\widehat{\ad\big(\op(b); \Psi_1\big)}(\bth, \bxi)= \sum_{\bchi\in\Bth\cup(\bth- \Bth)}\big(\hat{b}(\bth, \bxi)\hat{b}(\bth- \bchi, \bxi+ \bchi)\widehat{(\nabla_{\bchi}\tilde{\chi}_{\bth- \bchi})}(\bxi)\\ &+
\hat{b}(\bth, \bxi)\widehat{(\nabla_{\bchi}b)}(\bth- \bchi, \bxi)\tilde{\chi}_{\bth- \bchi}(\bxi)- \widehat{(\nabla_{\bth- \bchi}b)}(\bchi, \bxi)\hat b(\bth- \bchi, \bxi)\tilde\chi_{\bth- \bchi}(\bxi)\big),
\end{split}
\end{equation*}
\begin{equation*}%\label{ind2}
\begin{split}
\widehat{\ad(H_0; \Psi_1, \Psi_1)}(\bth, \bxi)&= \sum_{\bchi\in\Bth\cup(\bth- \Bth)}\big(\hat b(\bchi, \bxi+ \bth- \bchi)\widehat{(\nabla_{\bth- \bchi}\varphi_{\bchi} e_{\bchi})}(\bxi)\hat b(\bth- \bchi, \bxi)\tilde\chi_{\bth- \bchi}(\bxi)\\ &+ \widehat{(\nabla_{\bth- \bchi}b)}(\bchi, \bxi)\varphi_{\bchi}(\bxi)e_{\bchi}(\bxi)\hat b(\bth- \bchi, \bxi)\tilde\chi_{\bth- \bchi}(\bxi)\\ &- \varphi_{\bth}(\bxi)e_{\bth}(\bxi)\hat{b}(\bth, \bxi)\hat{b}(\bth- \bchi, \bxi+ \bchi)\widehat{(\nabla_{\bchi}\tilde{\chi}_{\bth- \bchi})}(\bxi)\\ &- \varphi_{\bth}(\bxi)e_{\bth}(\bxi)\hat{b}(\bth, \bxi)\widehat{(\nabla_{\bchi}b)}(\bth- \bchi, \bxi)\tilde{\chi}_{\bth- \bchi}(\bxi)\big).
\end{split}
\end{equation*}
Now, we complete the induction in several steps.

Step 1. First of all, notice that due to \eqref{bl:eq}, \eqref{tl:eq}, for any $m =2, \dots, \ell$ the symbol of $B_m$ admits a representation of the form \eqref{inds2} with $s= m- 1$, and symbol of $T_m$ admits a representation of the form \eqref{inds3} with $s= m$. Then it follows from Lemma~\ref{commut:lem} and \eqref{psil:eq} that $\Psi_\ell$ admits a representation of
the form \eqref{inds1}.

Step 2. Proof of \eqref{inds2} with $s =\ell$. Let $k_1 +\dots +k_p =\ell$. If $p \geqslant2$. Then
$$
\ad\big(\op(b); \Psi_{k_1}, \dots, \Psi_{k_p}\big)= \ad\Big(\ad\big(\op(b); \Psi_{k_1}, \dots, \Psi_{k_{p- 1}}\big); \Psi_{k_p}\Big).
$$
Since $k_1+ \dots+ k_{p- 1}\leqslant \ell- 1$ and $k_p\leqslant \ell- 1$ we can apply \eqref{inds1} and \eqref{inds2}. Combined with \eqref{comm:eq} it gives a representation of the form \eqref{inds2}. If $p= 1$ then $\ad\big(\op(b); \Psi_\ell\big)$ satisfies \eqref{inds2} because of \eqref{comm:eq} and step 1.

Step 3. Proof of \eqref{inds3} with $s= \ell+ 1$. Let $k_1+ \dots+ k_p= \ell+ 1$, $p\geqslant2$. If $p\geqslant3$, then (cf. step 2)
$$
\ad(H_0; \Psi_{k_1}, \dots, \Psi_{k_p})= \ad\big(\ad(H_0; \Psi_{k_1}, \dots, \Psi_{k_{p- 1}}); \Psi_{k_p}\big).
$$
Since $k_1+ \dots+ k_{p- 1}\leqslant \ell$, $p- 1\geqslant 2$ and $k_p\leqslant \ell- 1$ we can apply \eqref{inds1} and \eqref{inds3}. Together with \eqref{comm:eq} it gives a representation of the form \eqref{inds3}. If $p= 2$ then (see \eqref{psil:eq})
$$
\ad(H_0; \Psi_{k_1}, \Psi_{k_2}) =\ad\big(\ad(H_0; \Psi_{k_1}); \Psi_{k_2}\big) = -\ad(B_{k_1}^{\natural} +T_{k_1}^{\natural}; \Psi_{k_2}).
$$
Since $k_1 \leqslant\ell$ and $k_2 \leqslant\ell$, the representation of the form \eqref{inds3} follows from \eqref{comm:eq} and step 1. (Formally exceptional case $k_1 =1$, $k_2 =\ell$ can be treated separately in the same way using \eqref{psi1:eq} instead of \eqref{psil:eq}.)

Induction is complete.

Now, \eqref{inds2}, \eqref{inds3} and \eqref{Y_tilde k}, \eqref{bl:eq}, \eqref{tl:eq} prove the lemma.
\end{proof}

\section{Contribution from various resonant regions}\label{contribution section}

Let us fix a subspace $\GV \in\CV_m$, $m <d$, and a component $\Bxi_p$ of the resonant region $\Bxi(\GV)$.
Our aim is to compute the contribution to the density of states from each component $\Bxi_p$. Therefore, we define
\begin{equation}\label{Apm}
\CA^+_p(\rho):= \CA^+(\rho)\cap\Bxi_p\quad \textrm{and} \quad \CA^-_p(\rho):= \CA^-(\rho)\cap\Bxi_p
\end{equation}
and try to compute
\bee\label{eq:n3}
\vol\CA^+_p(\rho)- \vol\CA^-_p(\rho).
\ene
Since formulas \eqref{eq:46} and \eqref{eq:CARd}
obviously imply that
\bee\label{eq:n2}
\vol(\CG_{\la})= \omega_d\rho^d+ \sum_{m= 0}^{d -1}\sum_{\GV\in \CV_m}\sum_{p}\bigl(\vol\CA^+_p(\rho)- \vol\CA^-_p(\rho)\bigr),
\ene
Lemma~\ref{main_lem} would be proved if we manage to compute \eqref{eq:n3} (or at least prove that this expression admits a complete asymptotic expansion in $\rho$).

Note that if $\bxi\in\Bxi_p$, then we also have that $\BUps(\bxi)\subset \Bxi_p$. We denote
\bees
H_2(\bxi) :=H_2|_{\plainH{}_{\bxi}}, \quad \plainH{}_{\bxi} :=\CP\big(\BUps(\bxi)\big)\plainB_2(\R^d)
\enes
(recall that $\plainH{}_{\bxi}$ is an invariant subspace of $H_2$ acting in $\plainB_2(\R^d)$).
Suppose now that two points $\bxi$ and $\boldeta$ have the same coordinates $\mathbf X$ and $\mathbf\Phi$ and different coordinates $r$. Then $\bxi\in\Bxi_p$ implies $\boldeta \in\Bxi_p$ and $\BUps(\boldeta)= \BUps(\bxi)+ (\boldeta- \bxi)$. This shows that two spaces $\plainH{}_{\bxi}$ and $\plainH{}_{\boldeta}$ have the same dimension and, moreover,
there is a natural isometry $F_{\bxi,\boldeta}: \plainH{}_{\bxi}\to\plainH{}_{\boldeta}$ given by $F: \be_{\bnu}\mapsto\be_{\bnu+ (\boldeta- \bxi)}$, $\bnu \in\BUps(\bxi)$. This isometry allows us to `compare' operators acting in $\plainH{}_{\bxi}$ and
$\plainH{}_{\boldeta}$. Thus, abusing slightly our notation, we can assume that $H_2(\bxi)$ and $H_2(\boldeta)$
act in the same (finite dimensional) Hilbert space $\plainH{}(\mathbf X, \mathbf\Phi)$.
We will fix the values $(\mathbf X, \mathbf\Phi)$ and study how these operators depend on $r$. Thus, we denote by $H_2(r) =H_2(r; \mathbf X, \mathbf\Phi)$ the operator $H_2(\bxi)$ with $\bxi =(\mathbf X, r, \mathbf\Phi)$, acting in $\plainH{}(\mathbf X, \mathbf\Phi)$.

Let $W_{\tilde k}(r)$ be the operator in $\plainH{}(\mathbf X, \mathbf\Phi)$ with the symbol $w_{\tilde k}\big(\mathbf x,\boldsymbol\xi(\mathbf X, r, \mathbf\Phi)\big)$.
According to formula \eqref{a_dot_eta}, for any $s\leqslant \tilde k- 1$ and $\bth\in\Bth_{s+ 1}$
\begin{equation}\label{xi+ phi modulus squared}
|\bxi+ \bphi|^2= r^2+ 2r|\ba|\sum_{q= 1}^{K+ 1}a_{K+ 1\, q}\sin\Phi_q+ 2\lu\bxi, \bphi\ru+ |\mathbf X|^2+ |\ba|^2+ |\bphi|^2.
\end{equation}
This, together with \eqref{symbol series}, \eqref{b_iota} and \eqref{series for b}, implies that for $|\bxi+ \bphi|> C_0$ the coefficients $\hat{b}(\bth,\bxi+ \bphi)$ can be represented as the absolutely convergent series
\begin{equation}\label{b series}
\begin{split}
\hat{b}(\bth,\bxi+ \bphi)= \sum_{\iota\in \widetilde J}\sum_{l= 0}^\infty\sum_{\substack{n_1, \dots, n_{K+ 1}\geqslant 0\\ n_1+ \cdots+ n_{K+ 1}\leqslant l\\ j_1, \dots, j_d\geqslant 0\\ j_1+ \cdots+ j_d\leqslant l}}C^{\iota\, j_1\cdots j_d}_{l\, n_1\cdots n_{K+ 1}}(\mathbf X; \bth)r^{\iota- l}\phi_1^{j_1}\cdots\phi_d^{j_d}\prod_{a= 1}^{K+ 1}(\sin\Phi_a)^{n_a},
\end{split}
\end{equation}
where the coefficients satisfy
\begin{equation*}%\label{constants for b}
\big|C^{\iota\, j_1\cdots j_d}_{l\, n_1\cdots n_{K+ 1}}(\mathbf X; \bth)\big|\lesssim \rho_n^{(l- j_1- \cdots- j_d)(\al_{m+ 1}+ 0+)}
\end{equation*}

In the next lemma, to facilitate the expansion of the RHS of \eqref{symbol equation} in a suitable form, we transform the denominator of $\tilde\chi_{\bth'}$ (recall \eqref{chi}).

In the subsequent calculations we will use the generalized binomial coefficeints:
\begin{equation}\label{binomial coefficient}
\binom{p}{j}:= \begin{cases}
                1, & j= 0;\\ \displaystyle \frac1{j!}\prod_{k= 0}^{j- 1}(p- k), & j\in \N.
               \end{cases}
\end{equation}

\bel\label{denominator lemma}
For $s\leqslant \tilde k- 1$, $\bphi'\in\Bth_{s+ 1}$, $\bth'\in\Bth_{s+ 1}'$, and $\bxi$ in the support of $e_{\bth'}\varphi_{\bth'}$ let
\begin{equation*}
\begin{split}
D&:= \frac1w\sum_{j= 2}^\infty\binom{w}{j}r^{2- 2j}\sum_{k= 0}^{j- 1}\binom{j}{k}\Big(2r|\ba|\sum_{q= 1}^{K+ 1}a_{K+ 1\, q}\sin\Phi_q+ 2\lu\bxi, \bphi'\ru+ |\mathbf X|^2+ |\ba|^2+ |\bphi'|^2\Big)^{k}\\ &\times\big(2\lu\bxi, \bth'\ru+ 2\lu\bphi', \bth'\ru+ |\bth'|^2\big)^{j-k- 1}.
\end{split}
\end{equation*}
Then $|D|\lesssim \rho_n^{-1+ \al_{m+ 1}+ 0+}$ and
\begin{equation}\label{denominator formula}
\big(|\bxi+ \bphi'+ \bth'|^{2w}-|\bxi+ \bphi'|^{2w}\big)^{-1}= w^{-1}r^{2- 2w}\big(2\lu\bxi, \bth'\ru+ 2\lu\bphi', \bth'\ru+ |\bth'|^2\big)^{-1}\sum_{a= 0}^\infty (-D)^a.
\end{equation}
\enl

\bep
We introduce a shorthand
\begin{equation*}
N:= 2r|\ba|\sum_{q= 1}^{K+ 1}a_{K+ 1\, q}\sin\Phi_q+ 2\lu\bxi, \bphi'\ru+ |\mathbf X|^2+ |\ba|^2+ |\bphi'|^2.
\end{equation*}
Then by (generalized) binomial formula and \eqref{xi+ phi modulus squared} we obtain
\begin{equation}\label{D+ N}
\begin{split}
&|\bxi+ \bphi'+ \bth'|^{2w}-|\bxi+ \bphi'|^{2w}\\ &= \big(|\bxi|^2+ 2\lu\bxi, \bphi'+ \bth'\ru+ |\bphi'+ \bth'|^2\big)^w- \big(|\bxi|^2+ 2\lu\bxi, \bphi'\ru+ |\bphi'|^2\big)^w\\ &= \big(r^2+ N+ 2\lu\bxi, \bth'\ru+ 2\lu\bphi', \bth'\ru+ |\bth'|^2\big)^w- (r^2+ N)^w\\ &= r^{2w}\sum_{j= 1}^\infty\binom{w}{j}r^{-2j}\Big(\big(N+ 2\lu\bxi, \bth'\ru+ 2\lu\bphi', \bth'\ru+ |\bth'|^2\big)^j- N^j\Big)\\ &= wr^{2w- 2}\big(2\lu\bxi, \bth'\ru+ 2\lu\bphi', \bth'\ru+ |\bth'|^2\big)(1+ D).
\end{split}
\end{equation}
The estimate on $|D|$ follows from estimates \eqref{bound on R} and \eqref{bound on a}, and Lemmas~\ref{lem:Al} and \ref{lem:propBUps}. Now \eqref{denominator formula} follows from \eqref{D+ N}.
\enp

As we have seen from the previous sections, the symbol of the operator $H_2$ satisfies
\bee\label{eq:nn1}
h_2(\bx,\bxi)= |\bxi|^{2w}+ {w}_{\tilde k}(\bx,\bxi)= \big(r^2+ 2r\lu\ba, \mathbf\Phi\ru+ |\ba|^2+ |\mathbf X|^2\big)^w+ {w}_{\tilde k}(\bx,\bxi),
\ene
where $w_{\tilde k}$ are given by \eqref{eq:newy} and \eqref{symbol equation}.

\ber\label{cutoff remark}
In this section we assume that $\bxi\in\CA$, so by \eqref{bxi in CA} $2\rho_n/3\le |\bxi|\le 6\rho_n$, and by Remark~\ref{partition support remark} all functions $e_{\bth}(\bxi+ \cdot)$ from \eqref{eq:newy} and \eqref{symbol equation} are equal to $1$.
Note that if $\bth\in\Bth_{\tilde k}$, $\bphi\in\Bth_{\tilde k}$, and $\bth\not\in\GV$, then (see Lemma \ref{lem:products} and \eqref{phizeta:eq})
$\varphi_{\bth}(\bxi+ \bphi)= 1$. This means that all cut-off functions from \eqref{eq:newy} and \eqref{symbol equation} are equal to $1$ unless $\bth\in\GV$. If, on the other hand, $\bth\in\GV$, then $\varphi_{\bth}(\bxi+ \bphi)$
depends only on the projection $\bxi_{\GV}$ and thus is a function only of the coordinates $\mathbf X$.
\enr

By Proposition~\ref{bound:prop}, \eqref{eq:newy}, Lemma~\ref{symbol}, formulas \eqref{b series} and \eqref{denominator formula}, Lemma~\ref{lem:products}, and Remark~\ref{cutoff remark}, for $r\asymp \rho_n$
\begin{equation}\label{early derivative estimate}
\Big\|\frac{d^l}{dr^l}W_{\tilde k}(r)\Big\|\lesssim\rho_n^{\vark- l+ 0+}, \qquad l\geqslant 0.
\end{equation}

This, together with \eqref{eq:nn1}, implies
\bel\label{monotonicity of H_2(r) lemma}
The operator $H_2(r)$ is monotonically increasing in $r$; in particular, all its eigenvalues $\la_j\big(H_2(r)\big)$ are increasing in $r$.
\enl

Thus the function $g\big(\bxi(\mathbf X, r, \mathbf\Phi)\big)$ (defined in Section~\ref{description section}) is an increasing function of $r$ if we fix the other coordinates of $\bxi$, so the equation
\bees%\label{eq:tau}
g(\bxi)= \rho^{2w}
\enes
has a unique solution for fixed values of $\mathbf X$ and $\mathbf\Phi$; we
denote the $r$-coordinate of this solution by $\tau= \tau(\rho)= \tau(\rho; \mathbf X, \mathbf\Phi)$,
so that
\bee\label{eq:tau1}
g\big(\bxi(\mathbf X, \tau, \mathbf\Phi)\big)= \rho^{2w}.
\ene
By $\tau_0= \tau_0(\rho)= \tau_0(\rho; \mathbf X, \mathbf\Phi)$
we denote the value of $\tau$ for $(-\Delta)^w$, i.e. $\tau_0$ is a unique
solution of the equation
\bees
\big|\bxi(\mathbf X, \tau_0, \mathbf\Phi)\big|= \rho.
\enes
Obviously, we can write down a precise analytic expression for $\tau_0$ (and we have done this in \cite{ParSht} in the two-dimensional case) and show that it allows an expansion in powers of $\rho$ and $\ln\rho$, but we will not need it.
The definition \eqref{Apm} of the sets $\CA^{\pm}_p(\rho)$ implies that the intersection
\bees
\CA^+_p(\rho)\cap\big\{\bxi(\mathbf X, r, \mathbf\Phi),\ r\in\R_+\big\}
\enes
consists of points with $r$-coordinate belonging to the interval $\big[\tau_0(\rho), \tau(\rho)\big]$ (where we assume the interval to be empty if $\tau_0> \tau$). Similarly, the intersection
\bees
\CA^-_p(\rho)\cap\big\{\bxi(\mathbf X, r, \mathbf\Phi),\ r\in\R_+\big\}
\enes
consists of points with $r$-coordinate belonging to the interval $\big[\tau(\rho), \tau_0(\rho)\big]$.
Therefore,
\bees
\CA^+_p(\rho)= \Big\{\bxi= \bxi(\mathbf X, r, \mathbf\Phi), \mathbf X\in\Omega(\GV), \mathbf\Phi\in \CM_p, r\in\big[\tau_0(\rho; \mathbf X, \mathbf\Phi), \tau(\rho; \mathbf X, \mathbf\Phi)\big]\Big\}
\enes
and
\bees
\CA^-_p(\rho)= \Big\{\bxi= \bxi(\mathbf X, r, \mathbf\Phi), \mathbf X\in\Omega(\GV), \mathbf\Phi\in \CM_p, r\in\big[\tau(\rho; \mathbf X, \mathbf\Phi),\tau_0(\rho; \mathbf X, \mathbf\Phi)\big]\Big\}.
\enes
This implies that (recall that $K= d- m- 1$)
\bee\label{eq:n4}
\begin{split}
&\vol\CA^+_p(\rho)- \vol\CA^-_p(\rho) =\int_{\Omega(\GV)}d\mathbf X\int_{\CM_p}d\mathbf\Phi\int_{\tau_0(\rho; \mathbf X, \mathbf\Phi)}^{\tau(\rho; \mathbf X, \mathbf\Phi)}r^{K}dr\\
&= (K+ 1)^{-1}\int_{\CM_p}d\mathbf\Phi\int_{\Omega(\GV)}d\mathbf X\big(\tau(\rho; \mathbf X, \mathbf\Phi)^{K+ 1}- \tau_0(\rho; \mathbf X, \mathbf\Phi)^{K+ 1}\big).
\end{split}
\ene

\ber\label{K= 0 remark}
Note that in the case $K= 0$ the simplex $\CM_p$ is degenerate and there is no integration in $d\mathbf\Phi$.
\enr

Obviously, it is enough to compute the part of \eqref{eq:n4} containing $\tau$, since the second part (containing $\tau_0$) can be computed analogously. We start by considering
\bee\label{eq:n5}
\int_{\Omega(\GV)}\tau(\rho; \mathbf X, \mathbf\Phi)^{K+ 1}d\mathbf X.
\ene
First of all, we notice that if $\bxi,\boldeta\in\BXi(\GV)$ are resonant congruent points then, according to Lemma~\ref{lem:Upsilon}, all vectors $\bth_j$ from Definition~\ref{reachability:defn} of equivalence belong to $\GV$. This naturally leads to the definition of equivalence for projections $\bxi_{\GV}$ and $\boldeta_{\GV}$. Namely, we say that two points $\bnu$ and $\bmu$ from $\Omega(\GV)$ are $\GV$-equivalent (and write $\bnu\leftrightarrow_{\GV}\bmu$) if $\bnu$ and $\bmu$ are equivalent in the sense of Definition~\ref{reachability:defn} with an additional requirement that all $\bth_j\in\GV$. Then
$\bxi \leftrightarrow\boldeta$ implies $\bxi_{\GV} \leftrightarrow_{\GV}\boldeta_{\GV}$. For $\bnu \in\Omega(\GV)$ we denote by $\BUps_{\GV}(\bnu)$ the class of equivalence of $\bnu$ generated by $\leftrightarrow_{\GV}$. Then $\BUps_{\GV}(\bxi_{\GV})$ is a projection of $\BUps(\bxi)$ to $\GV$ and is, therefore, finite.

Since $\BUps_{\GV}(\bnu)$ is a finite set for each $\bnu\in\Omega(\GV)$, we can re-write \eqref{eq:n5} as
\bee\label{eq:n6}
\int_{\Omega(\GV)}\tau(\rho; \mathbf X, \mathbf\Phi)^{K+ 1}d\mathbf X= \int_{\Omega(\GV)}\big(\card\BUps_{\GV}(\bnu)\big)^{-1}\sum_{\mathbf X\in\BUps_{\GV}(\bnu)}\tau(\rho; \mathbf X, \mathbf\Phi)^{K+ 1}d\bnu
\ene
and try to compute
\bees%\label{eq:n7}
\sum_{\mathbf X\in\BUps_{\GV}(\bnu)}\tau(\rho; \mathbf X, \mathbf\Phi)^{K+ 1}.
\enes

Remark~\ref{cutoff remark}, together with equations \eqref{eq:nn1}, \eqref{eq:newy}, and \eqref{symbol equation}, shows that $H_2(r)$ depends on $r$ analytically, so we can and will consider the family $H_2(z)$ with complex values of the parameter $z$ with $\Re e~z\asymp \rho$. Likewise, we analytically continue the function $\bxi(\mathbf X, r, \mathbf\Phi)$ to
\begin{equation}\label{extended xi}
\bxi(\mathbf X, z, \mathbf\Phi):= \mathbf X+ \ba+ z\mathbf\Phi.
\end{equation}
We also introduce the analytic continuation $|\cdot|_{\C}$ of the modulus of vectors, so that
\bee\label{modulus complexified}
|\bxi|_{\C}^{2}:= z^2+ 2z\lu\ba, \mathbf\Phi\ru+ |\ba|^2+ |\mathbf X|^2.
\ene
Formulas \eqref{eq:newy} and \eqref{symbol equation} give matrix elements of $H_2(z)$ in an orthonormal basis even for complex $z$.

We choose a contour
\begin{equation}\label{contour}
\g:= \bigg\{z\in\mathbb C:\, |z- \rho|= t\rho_n:= \Big(8\max\big\{(2w- 2)/3, 1\big\}\Big)^{-1}\rho_n\bigg\}
\end{equation}
to be a circle in the complex plane going in the positive direction.

Estimates \eqref{early derivative estimate} remain valid after the analytic continuation: for all $z$ inside and on $\g$
\begin{equation}\label{derivative estimate}
\Big\|\frac{d^l}{dz^l}W_{\tilde k}(z)\Big\|\lesssim\rho_n^{\vark- l+ 0+}, \qquad l\geqslant 0.
\end{equation}

\bel\label{key_lemma}
For $\rho\in I_n= [\rho_n, 4\rho_n]$ all $\tau(\rho; \mathbf X, \mathbf\Phi)$ lie inside $\gamma$. These are the only zeros of the function $\det\big(H_2(z)- \rho^{2w}I\big)$ inside the contour.
\enl

\bep
Let $r:= \Re e~z$, $y:= \Im m~z$.
For $y= 0$ the operator $H_2(r)$ is self-adjoint. Thus it has $\card\BUps_{\mathfrak V}(\boldsymbol\nu)$ real eigenvalues.

Now for $r\geqslant \rho+ t\rho_n\geqslant (1+ t/4)\rho$ relations \eqref{eq:nn1}, \eqref{bound on a}, Lemma~\ref{lem:propBUps}(ii), and \eqref{derivative estimate} imply
\begin{equation*}
 H_2(r)\geqslant \Big(\big((1+ t/4)\rho\big)^{2w}\big(1- O(\rho^{\al_{m+ 1}- 1+ 0+})\big)- O(\rho^{\vark+ 0+})\Big)I.
\end{equation*}
Thus by \eqref{beta and alphas} and \eqref{alpha} for big $\rho$ no eigenvalue of $H_2(r)$ can coincide with $\rho^{2w}$.

Likewise for $r\leqslant \rho- t\rho_n\leqslant (1- t/4)\rho$ for big $\rho$ we have
\begin{equation*}
 H_2(r)\leqslant \Big(\big((1- t/4)\rho\big)^{2w}\big(1+ O(\rho^{\al_{m+ 1}- 1+ 0+})\big)+ O(\rho^{\vark+ 0+})\Big)I,
\end{equation*}
and no eigenvalue of $H_2(r)$ can coincide with $\rho^{2w}$. This implies that all the eigenvalues of $H_2(r)$ lie in the real interval $(\rho- t\rho_n, \rho+ t\rho_n)$. By \eqref{eq:tau1} and Lemma~\ref{monotonicity of H_2(r) lemma} these eigenvalues coincide with $\big\{\tau(\rho; \mathbf X, \mathbf\Phi): \mathbf X\in\BUps_{\GV}(\bnu)\big\}$.

It remains to show that $H_2(r+ iy)$ is invertible for any nonzero $y$ such that $r+ iy$ is inside or on $\g$.
Relation \eqref{extended xi}, Lemma~\ref{lem:propBUps}(ii), definition \eqref{L_j}, and bound \eqref{bound on a} imply that inside and on the contour
\begin{equation*}%\label{essence of xi}
\bxi= (r+ iy)\big(1+ O(\rho_n^{-1+ \alpha_{m+ 1}+ 0+})\big)
\end{equation*}
and
\begin{equation*}%\label{estimate on argument}
\mathrm{arg\,}|\boldsymbol\xi|_{\C}\leqslant \big(1+ o(1)\big)\arcsin(t\rho_n/\rho)\leqslant \big(1+ o(1)\big)\arcsin t\leqslant t\big(1+ o(1)\big).
\end{equation*}
Hence
\begin{equation*}%\label{complex xi estimate}
\big||\boldsymbol\xi|_{\C}^{2w}\big|= \big||\boldsymbol\xi|_{\C}^2\big|^w\asymp \rho^{2w}\quad \textrm{and}\quad \mathrm{arg\,}|\boldsymbol\xi|_{\C}^{2w}= w\arcsin\frac{2y\big(r+ \langle\mathbf a, \mathbf\Phi\rangle\big)}{\big||\boldsymbol\xi|_{\C}^2\big|}\asymp y\rho^{-1},
\end{equation*}
which implies that
\begin{equation}\label{Im of main symbol}
\Big|\mathrm{Im}\big(|\boldsymbol\xi|_{\C}^{2w}\big)\Big|\gtrsim |y|\rho^{2w- 1}.
\end{equation}
Now for any $\Psi\in\plainH{}(\mathbf X, \mathbf\Phi)$ with $\|\Psi\|= 1$ we have by \eqref{Im of main symbol} and \eqref{derivative estimate}
\begin{equation*}\label{idea}
\begin{split}
\Big\|\big(H_2(z)- \rho^{2w}I\big)\Psi\Big\|&\geqslant \Big|\textrm{Im}\langle\big(H_2(z)- \rho^{2w}I\big)\Psi, \Psi\rangle\Big|\\ &\geqslant \Big|\textrm{Im}\big(|\boldsymbol\xi|_{\C}^{2w}\big)\Big|- |y|\underset{t\in[0, y]}{\textrm{sup}}\big\|W'(r+ it)\big\|\gtrsim |y|\rho^{2w- 1},
\end{split}
\end{equation*}
where we have used that for $y= 0$ the quadratic form of $W(z)$ is real-valued. So the kernel of $H_2(r+ iy)- \rho^{2w}$ is trivial for $y\neq 0$.
\enp

\begin{lem}\label{z-denominator lemma}
 For $z\in\g$ and $l\in \N$
 \begin{equation}\label{denominator representation}
  (z^{2w}- \rho^{2w})^{-l}= \rho^{-2wl}\sum_{j= 0}^\infty A_{l\, j}\Big(\frac{z- \rho}\rho\Big)^{j- l},
 \end{equation}
 where
 \begin{equation*}
  A_{l\, j}=\begin{cases}
       (2w)^{-l}, & j= 0;\\
       \displaystyle\frac1{(2w)^l}\sum_{p= 1}^j\frac1{(2w)^{p}}\binom{-l}{p}\sum_{\substack{q_1, \dots, q_p\geqslant 1\\ q_1+ \cdots +q_p= j}}\binom{2w}{q_1+ 1}\binom{2w}{q_2+ 1}\cdots\binom{2w}{q_p+ 1}, & j> 0.
      \end{cases}
 \end{equation*}
 The series in \eqref{denominator representation} converges absolutely.
\end{lem}

\bep
A striaghtforward calculation gives
\begin{equation}\label{before progression}
\begin{split}
 (z^{2w}- \rho^{2w})^{-l}&= \frac1{\rho^{2wl}}\bigg(\Big(1+ \frac{z- \rho}{\rho}\Big)^{2w}- 1\bigg)^{-l}= \frac1{\rho^{2wl}}\bigg(\sum_{q= 1}^\infty\binom{2w}{q}\Big(\frac{z- \rho}{\rho}\Big)^q\bigg)^{-l}\\ &= \frac{\rho^{-2wl}}{(2w)^l}\Big(\frac{z- \rho}{\rho}\Big)^{-l}\bigg(1+ \frac1{2w}\sum_{q= 1}^\infty\binom{2w}{q+ 1}\Big(\frac{z- \rho}{\rho}\Big)^q\bigg)^{-l}.
\end{split}
\end{equation}
If $2w\in\mathbb N$, then the series on the right hand side is finite. Otherwise, by \eqref{contour} and \eqref{binomial coefficient}, for $z\in\g$ the ratio of absolute values of any two sequential terms of the series satisfies
\begin{equation*}
\bigg|\frac{z-\rho}\rho\binom{2w}{q+ 2}\binom{2w}{q+ 1}^{-1}\bigg|= \Big|\frac{z-\rho}\rho\Big|\frac{|2w- q- 1|}{q+ 2}\leqslant \frac18, \qquad q\geqslant 1.
\end{equation*}
So, again by \eqref{contour} and \eqref{binomial coefficient}, we have
\begin{equation*}
\bigg|\sum_{q= 1}^\infty\binom{2w}{q+ 1}\Big(\frac{z- \rho}{\rho}\Big)^q\bigg|< \bigg|\binom{2w}{2}\bigg|\frac{|z- \rho|}\rho\sum_{q= 0}^\infty\frac1{8^q}\leqslant \frac{4w}7.
\end{equation*}
Thus we can decompose the expression on the right hand side of \eqref{before progression} into an absolutely converging series obtaining
\begin{equation*}
\begin{split}
&(z^{2w}- \rho^{2w})^{-l}= \frac{\rho^{-2wl}}{(2w)^l}\Big(\frac{z- \rho}{\rho}\Big)^{-l}\sum_{p= 0}^\infty\binom{-l}{p}\frac1{(2w)^p}\bigg(\sum_{q= 1}^\infty\binom{2w}{q+ 1}\Big(\frac{z- \rho}{\rho}\Big)^q\bigg)^p\\ &= \frac{\rho^{-2wl}}{(2w)^l}\Big(\frac{z- \rho}{\rho}\Big)^{-l}\bigg(1+ \sum_{j= 1}^\infty\Big(\frac{z- \rho}{\rho}\Big)^{j}\sum_{p= 1}^j\frac1{(2w)^p}\binom{-l}{p}\sum_{\substack{q_1, \dots, q_p\geqslant 1\\ q_1+ \cdots +q_p= j}}\binom{2w}{q_1+ 1}\cdots\binom{2w}{q_p+ 1}\bigg),
\end{split}
\end{equation*}
which finishes the proof.
\enp

Let $S(z):= H_2(z)- z^{2w}I$ in $\plainH{}(\mathbf X, \mathbf\Phi)$. Then by \eqref{eq:nn1} on $\g$ the symbol of $S(z)$ admits the representaion
\begin{equation}\label{symbol for S}
 s(z)= \sum_{v= 1}^\infty\binom{w}{v}z^{2w- v}\Big(2\lu\ba, \mathbf\Phi\ru+ z^{-1}\big(|\ba|^2+|\mathbf X|^2\big)\Big)^v+ w_{\tilde k}(z).
\end{equation}

Relations \eqref{symbol for S}, \eqref{derivative estimate}, \eqref{bound on a}, Lemma~\ref{lem:propBUps}(ii), and \eqref{beta and alphas} imply that everywhere inside and on $\g$
\bee\label{eq:newS2}
\Big\|\frac{d^l}{dz^l}S(z)\Big\|\lesssim \rho_n^{2w- 1+ \al_{m+ 1}- l+ 0+},\ \ l\ge 0.
\ene

A version of the Jacobi's formula states that for any differentiable invertible matrix-valued function $F(z)$ we have
$$
\tr\big[F'(z)F^{-1}(z)\big]=\Big(\det\big[F(z)\big]\Big)'\Big(\det\big[F(z)\big]\Big)^{-1}
$$
(it can be proved, for example, using the expansion of the determinant along rows and the induction in the size of $F$).

Then by Lemma~\ref{z-denominator lemma} and the residue theorem
\bee\label{eq:residues}
\begin{split}
&\sum_{\mathbf X\in\BUps_{\GV}(\bnu)}\tau(\rho;\mathbf X,\mathbf\Phi)^{K+ 1}\\
&= \frac{1}{2\pi i}\oint_\gamma z^{K+ 1}\Big(\det\big[H_2(z)- \rho^{2w}I\big]\Big)'\Big(\det\big[H_2(z)- \rho^{2w}I\big]\Big)^{-1}dz\\
&= \frac{1}{2\pi i}\oint_\gamma \tr\Big[z^{K+ 1} H_2'(z)\big(H_2(z)- \rho^{2w}I\big)^{-1}\Big]dz\\
&= \frac{1}{2\pi i}\oint_\gamma \tr\Big[\big(2wz^{2w+ K}I+ z^{K+ 1}S'(z)\big)\sum_{l=0}^\infty (-1)^lS^l(z)(z^{2w}- \rho^{2w})^{-1- l}\Big]dz\\
&= \frac{1}{2\pi i}\oint_\gamma \tr\Big[\big(2wz^{2w+ K}I+ z^{K+ 1}S'(z)\big)\\ &\qquad\times\sum_{l= -\infty}^\infty(z- \rho)^{-1- l}\sum_{j= 0}^\infty(-1)^{l+ j}A_{1+ l+ j\, j}\rho^{1+ l- 2w(1+ l+ j)}S^{l+ j}(z)\Big]dz\\
&= \sum_{l= 0}^\infty\frac1{l!}\tr\frac{d^l}{dr^l}\Big[\big(2wr^{2w+ K}I+ r^{K+ 1}S'(r)\big)\sum_{j= 0}^\infty(-1)^{l+ j}A_{1+ l+ j\, j}\rho^{1+ l- 2w(1+ l+ j)}S^{l+ j}(r)\Big]\Big|_{r= \rho}.
\end{split}
\ene

We can restrict the summation on the RHS of \eqref{eq:residues} to
\begin{equation*}%\label{l_0}
l+ j\leqslant l_0:= \big(M+ K+ d+ 1+ (d- 1)\al_{d- 1}- 2w\big)/(1- \al_{m+ 1}).
\end{equation*}
Indeed, using the trivial fact that for any linear operator $A$ in the finite dimensional Hilbert space spanned by $\be_{\bth}$ with $\bth\in \BUps_{\GV}(\bnu)$
\begin{equation*}%\label{norm to trace}
|\tr A|\leqslant \|A\|\card\BUps_{\GV}(\bnu),
\end{equation*}
estimate \eqref{eq:newS2}, and relation \eqref{beta and alphas} we can see that the sum of the terms in \eqref{eq:residues} with $l+ j> l_0$ contributes only to the order $O(\rho_n^{-M+ 2w- d})$ in \eqref{eq:n6}, and thus after integration in $\mathbf\Phi$ the corresponding term can be included into the remainder $R_{\tilde k}$ of Section~\ref{description section}.

Formula \eqref{eq:n4} shows that in order to compute the contribution to the density of states
from $\Bxi(\GV)_p$, we need to integrate the RHS of \eqref{eq:residues} against $d\bnu$ and $d\mathbf\Phi$. We are going to integrate against $d\mathbf\Phi$ first:
\begin{equation}\label{integration in Phi}
\begin{split}
&\int_{\CM_p}d\mathbf\Phi\int_{\Omega(\GV)}\big(\card\BUps_{\GV}(\bnu)\big)^{-1}\sum_{\mathbf
X\in\BUps_{\GV}(\bnu)}\tau(\rho; \mathbf X, \mathbf\Phi)^{K+ 1}\\ &=
\int_{\Omega(\GV)}d\bnu\big(\card\BUps_{\GV}(\bnu)\big)^{-1}\sum_{l=
0}^{l_0}\sum_{j= 0}^{l_0- l}\frac{(-1)^{l+ j}}{l!}A_{1+ l+ j\,
j}\rho^{1+ l- 2w(1+ l+ j)}\\ &\times
\tr\frac{d^l}{dr^l}\Big[\int_{\CM_p}d\mathbf\Phi\big(2wr^{2w+ K}I+
r^{K+ 1}S'(r)\big)S^{l+ j}(r)\Big]\Big|_{r= \rho}+ O(\rho_n^{-M+ 2w-
d})\\ &= O(\rho_n^{-M+ 2w- d})+
\int_{\Omega(\GV)}\frac{d\bnu}{\card\BUps_{\GV}(\bnu)}\sum_{l=
0}^{l_0}\sum_{j= 0}^{l_0- l}\frac{(-1)^{l+ j}}{l!}A_{1+ l+ j\, j}
\rho^{1+ l- 2w(1+ l+ j)}\\
&\times\tr\bigg[\frac{d^l}{dr^l}\Big(2wr^{2w+ K}\int_{\CM_p} S^{l+
j}(r)d\mathbf\Phi - \frac{(K+1)r^{K}}{l+ j+ 1}\int_{\CM_p}S^{l+ j+
1}(r)d\mathbf\Phi\Big)\\ &+ \frac{d^{l+ 1}}{dr^{l+
1}}\Big(\frac{r^{K+ 1}}{l+ j+ 1}\int_{\CM_p}S^{l+ j+
1}(r)d\mathbf\Phi\Big)\bigg]\bigg|_{r= \rho}.
\end{split}
\end{equation}

We will prove that the integrand of the exterior integral in \eqref{integration in Phi} is a convergent series of products of powers of $\rho$ and $\ln\rho$. The coefficients in front of all terms will be bounded functions
of $\mathbf X$, so afterwards we will just integrate these coefficients to obtain the desired asymptotic expansion.

Let us discuss, how $S(r)$ depends on $\rho$, $\mathbf X$ and $\mathbf\Phi$.
In order to do this, we first look again at \eqref{symbol equation}. As follows from Remark~\ref{cutoff remark}, the product $e_{\bth_q''}\varphi_{\bth_q''}$ does not depend on $r$ and $\mathbf\Phi$, and by \eqref{varphi:eq}
\begin{equation}\label{Delta cutoffs}
\|\widehat{\nabla^{\nu}e_{\bth_q''}\varphi_{\bth_q''}}\|_{\plainL\infty(\R^d)}\lesssim \rho_n^{-\nu\b}.
\end{equation}
For any $\boldeta\in\Bth_{s+ 1}$ the application of the finite difference operator $\nabla_{\boldeta}$ to a polynomial decreases its degree by $1$. Hence formula \eqref{b series} ensures that
\begin{equation}\label{Delta b}
\begin{split}
\widehat{(\nabla^{\nu}b)}(\bth, \bxi+ \bphi)= \sum_{\iota\in \widetilde J}\sum_{i= \nu}^\infty\sum_{\substack{n_1, \dots, n_{K+ 1}\geqslant 0\\ n_1+ \cdots+ n_{K+ 1}\leqslant i\\ j_1, \dots, j_d\geqslant 0\\ j_1+ \cdots+ j_d\leqslant i- \nu}}\widetilde C^{\iota\, j_1\cdots j_d}_{i\, n_1\cdots n_{K+ 1}}(\mathbf X; \bth)r^{\iota- i}\phi_1^{j_1}\cdots\phi_d^{j_d}\prod_{a= 1}^{K+ 1}(\sin\Phi_a)^{n_a}.
\end{split}
\end{equation}
Here $\widetilde C^{\iota\, j_1\cdots j_d}_{i\, n_1\cdots n_{K+ 1}}(\mathbf X; \bth)$ depend on the coefficients of \eqref{Delta-power} and satisfy a uniform estimate
\begin{equation*}%\label{Delta b coefficients}
\big|\widetilde C^{\iota\, j_1\cdots j_d}_{i\, n_1\cdots n_{K+ 1}}(\mathbf X; \bth)\big|\lesssim \rho_n^{(i- \nu- j_1- \cdots- j_d)(\al_{m+ 1}+ 0+)}.
\end{equation*}
Now
\begin{equation*}%\label{Delta chi}
\widehat{(\nabla^{\nu}\tilde\chi_{\bth})}(\bxi)= \sum_{\tilde\nu= 0}^{\nu}\widehat{(\nabla^{\tilde\nu}e_{\bth}\varphi_{\bth})}\Big(\bxi+ \sum_{p= 1}^{\tilde\nu}\boldeta_k\Big)\widehat{\Big(\nabla^{\nu- \tilde\nu}\big(|\cdot+ \bth|_{\C}^{2w}- |\cdot|_{\C}^{2w}\big)^{-1}\Big)}(\bxi).
\end{equation*}
The factors $\widehat{(\nabla^{\tilde\nu}e_{\bth}\varphi_{\bth})}$ satisfy the estimate \eqref{Delta cutoffs}. For $\boldeta\in \Bth_{s+ 1}$ we have
\begin{equation*}%\label{Once different denominator}
\begin{split}
&\widehat{\Big(\nabla_{\boldeta}\big(|\cdot+ \bth|_{\C}^{2w}- |\cdot|_{\C}^{2w}\big)^{-1}\Big)}(\bxi)\\ &= \big(|\bxi+ \boldeta+ \bth|_{\C}^{2w}- |\bxi+ \boldeta|_{\C}^{2w}\big)^{-1}\big(|\bxi+ \bth|_{\C}^{2w}- |\bxi|_{\C}^{2w}\big)^{-1}G(\bxi; \bth, \boldeta),
\end{split}
\end{equation*}
where
\begin{equation*}%\label{G}
\begin{split}
&G(\bxi; \bth, \boldeta):= |\bxi+ \bth|_{\C}^{2w}- |\bxi|_{\C}^{2w}- |\bxi+ \boldeta+ \bth|_{\C}^{2w}+ |\bxi+ \boldeta|_{\C}^{2w}\\ &= -2w\lu\boldeta, \bth\ru|\bxi|_{\C}^{2w- 2}\\ &+ \sum_{j= 2}^\infty\binom wj|\bxi|_{\C}^{2w- 2j}\Big(\big(2\lu\bxi, \bth\ru+ |\bth|^2\big)^j- \big(2\lu\bxi, \boldeta+ \bth\ru+ |\boldeta+ \bth|^2\big)^j+ \big(2\lu\bxi, \boldeta\ru+ |\boldeta|^2\big)^j\Big).
\end{split}
\end{equation*}
In analogy to \eqref{Delta b} we have
\begin{equation*}%\label{Delta G}
\begin{split}
\widehat{\big(\nabla^{\nu}G(\cdot; \bth, \boldeta)\big)}(\bxi)= \sum_{i= \nu}^\infty\sum_{\substack{n_1, \dots, n_{K+ 1}\geqslant 0\\ n_1+ \cdots+ n_{K+ 1}\leqslant i+ 2}}\widetilde C^i_{n_1\cdots n_{K+ 1}}(\mathbf X; \bth, \boldeta)r^{2w- 2- i}\prod_{a= 1}^{K+ 1}(\sin\Phi_a)^{n_a},
\end{split}
\end{equation*}
with
\begin{equation}\label{Delta G coefficients}
\big|\widetilde C^i_{n_1\cdots n_{K+ 1}}(\mathbf X; \bth, \boldeta)\big|\lesssim \rho_n^{(i- \nu)(\al_{m+ 1}+ 0+)+ 0+}.
\end{equation}
Altogether, applying relations \eqref{Delta cutoffs} -- \eqref{Delta G coefficients} to \eqref{eq:newy} and \eqref{symbol equation} we obtain
\begin{equation}\label{W series}
\begin{split}
&w_{\tilde k}(\bth, \bxi)\\ &= \sum_{s= 0}^{\tilde k- 1}\sum_{\iota_0, \dots, \iota_s\in \widetilde J}\sum_{\mu= 0}^s\sum_{\substack{\boldeta_1, \dots, \boldeta_{s+ \mu}\in \Bth_{s+ 1}\\ \bth_1, \dots, \bth_{s+ \mu}\in \Bth'_{s+ 1}}}\sum_{p= 0}^{s- \mu}\sum_{i= 0}^\infty\sum_{\substack{n_1, \dots, n_{K+ 1}\geqslant 0\\ n_1+ \cdots+ n_{K+ 1}\leqslant 2\mu+ p+ i}} C_{s\, \mu\, p\, i\, \iota_0\cdots \iota_s\, n_1\cdots n_{K+ 1}}^{\boldeta_1\cdots \boldeta_{s+ \mu}\, \bth_1\cdots \bth_{s+ \mu}}(\mathbf X; \bth)\\ &\times r^{(2w- 2)\mu+ \iota_0+ \cdots+ \iota_s- p- i}\prod_{a= 1}^{K+ 1}(\sin\Phi_a)^{n_a}\prod_{v= 1}^{s+ \mu}\big(|\bxi+ \boldeta_v+ \bth_v|_{\C}^{2w}- |\bxi+ \boldeta_v|_{\C}^{2w}\big)^{-1},
\end{split}
\end{equation}
where
\begin{equation*}%\label{W series coefficients}
\big|C_{s\, \mu\, p\, i\, \iota_0\cdots \iota_s\, n_1\cdots n_{K+ 1}}^{\boldeta_1\cdots \boldeta_{s+ \mu}\, \bth_1\cdots \bth_{s+ \mu}}(\mathbf X; \bth)\big|\lesssim \rho_n^{i(\al_{m+ 1}+ 0+)- (s- \mu- p)\b+ 0+}.
\end{equation*}
According to Lemma~\ref{denominator lemma},
\begin{equation}\label{old to new denominator}
\begin{split}
&\big(|\bxi+ \boldeta_v+ \bth_v|_{\C}^{2w}- |\bxi+ \boldeta_v|_{\C}^{2w}\big)^{-1}\\ &= r^{2- 2w}\big(2\lu\bxi, \bth_v\ru+ 2\lu\boldeta_v, \bth_v\ru+ |\bth_v|^2\big)^{-1}\\ &\times\sum_{i= 0}^\infty \sum_{\substack{n_1, \dots, n_{K+ 1}\geqslant 0\\ n_1+ \cdots+ n_{K+ 1}\leqslant i}}C_{n_1\cdots n_{K+ 1}}^i(\mathbf X; \boldeta_v, \bth_v)r^{-i}\prod_{a= 1}^{K+ 1}(\sin\Phi_a)^{n_a},
\end{split}
\end{equation}
and here
\begin{equation*}%\label{denominator coefficients}
\big|C_{n_1\cdots n_{K+ 1}}^i(\mathbf X; \boldeta_v, \bth_v)\big|\lesssim \rho_n^{i(\al_{m+ 1}+ 0+)}.
\end{equation*}
If now subsitute \eqref{old to new denominator} to \eqref{W series}, we obtain
\begin{equation}\label{new W series}
\begin{split}
&w_{\tilde k}(\bth, \bxi)\\ &= \sum_{s= 0}^{\tilde k- 1}\sum_{\iota_0, \dots, \iota_s\in \widetilde J}\sum_{\mu= 0}^s\sum_{\substack{\boldeta_1, \dots, \boldeta_{s+ \mu}\in \Bth_{s+ 1}\\ \bth_1, \dots, \bth_{s+ \mu}\in \Bth'_{s+ 1}}}\sum_{p= 0}^{s- \mu}\sum_{i= 0}^\infty\sum_{\substack{n_1, \dots, n_{K+ 1}\geqslant 0\\ n_1+ \cdots+ n_{K+ 1}\\ \leqslant 2\mu+ p+ i}} C_{s\, \mu\, p\, i\, \iota_0\cdots \iota_s\, n_1\cdots n_{K+ 1}}^{\boldeta_1\cdots \boldeta_{s+ \mu}\, \bth_1\cdots \bth_{s+ \mu}}(\mathbf X; \bth, \boldeta_v, \bth_v)\\ &\times r^{(2- 2w)s+ \iota_0+ \cdots+ \iota_s- p- i}\prod_{a= 1}^{K+ 1}(\sin\Phi_a)^{n_a}\prod_{v= 1}^{s+ \mu}\big(2\lu\bxi, \bth_v\ru+ 2\lu\boldeta_v, \bth_v\ru+ |\bth_v|^2\big)^{-1},
\end{split}
\end{equation}
with
\begin{equation*}%\label{new W coefficients}
\big|C_{s\, \mu\, p\, i\, \iota_0\cdots \iota_s\, n_1\cdots n_{K+ 1}}^{\boldeta_1\cdots \boldeta_{s+ \mu}\, \bth_1\cdots \bth_{s+ \mu}}(\mathbf X; \bth, \boldeta_v, \bth_v)\big|\lesssim \rho_n^{i(\al_{m+ 1}+ 0+)- (s- \mu- p)\b+ 0+}.
\end{equation*}

The first sum in \eqref{symbol for S} can be written in the form
\begin{equation}\label{first sum for S}
\begin{split}
&\sum_{v= 1}^\infty\binom{w}{v}z^{2w- v}\Big(2\lu\ba, \mathbf\Phi\ru+ z^{-1}\big(|\ba|^2+ |\mathbf X|^2\big)\Big)^v\\ &= \sum_{i= 0}^\infty\sum_{\substack{n_1, \dots, n_{K+ 1}\geqslant 0\\ n_1+ \cdots+ n_{K+ 1}\leqslant i+ 1}}C^i_{n_1\cdots n_{K+ 1}}(\mathbf X)z^{2w- 1- i}\prod_{a= 1}^{K+ 1}(\sin\Phi_a)^{n_a},
\end{split}
\end{equation}
where
\begin{equation*}%\label{first sum coefficients}
\big|C^i_{n_1\cdots n_{K+ 1}}(\mathbf X)\big|\lesssim \rho_n^{(i+ 1)(\al_{m+ 1}+ 0+)}.
\end{equation*}

Substituting \eqref{new W series} and \eqref{first sum for S} into \eqref{symbol for S} we can calculate the series for the symbol of the operator $S^f$ for $f\in \N$:
\begin{equation}\label{S^f}
\begin{split}
&\widehat{s^f}(\bth, \bxi)= \sum_{\substack{\bth_1, \dots, \bth_f\in \Bth_{\tilde k}\\ \bphi_1, \dots, \bphi_{f}\in \Bth_{\tilde k}}}C_{\bphi_1\cdots \bphi_{f}}^{\bth_1\cdots \bth_{f}}(\bth)\prod_{g= 1}^f\hat s(\bth_g, \bxi+ \bphi_g)\\ &= \sum_{\nu= 0}^{f}\sum_{h= \nu}^{\nu\tilde k}\sum_{\iota_1, \dots, \iota_h\in \widetilde J}\sum_{\mu= 0}^{h- \nu}\sum_{\substack{\bphi_1, \dots, \bphi_{h- \nu+ \mu}\in \Bth_{\tilde k}\\ \bth_1, \dots, \bth_{h- \nu+ \mu}\in \Bth'_{\tilde k}}}\sum_{p= 0}^{h- \nu- \mu}\sum_{i= 0}^\infty\sum_{\substack{n_1, \dots, n_{K+ 1}\geqslant 0\\ n_1+ \cdots+ n_{K+ 1}\leqslant 2\mu+ p+ i+ f- \nu}}C(\mathbf X; \bth, \dots)\\ &\times r^{(2- 2w)h+ (2w- 1)f- \nu+ \iota_1+ \cdots+ \iota_h- p- i}\prod_{a= 1}^{K+ 1}(\sin\Phi_a)^{n_a}\prod_{v= 1}^{h- \nu+ \mu}\big(2\lu\bxi, \bth_v\ru+ 2\lu\bphi_v, \bth_v\ru+ |\bth_v|^2\big)^{-1},
\end{split}
\end{equation}
with
\begin{equation*}%\label{S^f constant}
\big|C(\mathbf X; \bth, \dots)\big|\lesssim \rho_n^{(f- \nu+ i)(\al_{m+ 1}+ 0+)- (h- \nu- \mu- p)\b+ 0+}.
\end{equation*}

Note that the last product on the right hand side of \eqref{S^f} is of the form
\bees%\label{eq:f3}
\prod_{t =1}^T\Big(l_t +\rho\sum_{q} b_q^t\sin \Phi_q\Big)^{-k_t}.
\enes
Here we have expanded the inner products $\lu\bxi,\bth_v\ru$ using Lemma \ref{lem:products}(ii).
The coefficients $\{b_q^t\}$ in the decomposition $(\bth_v)_{\GV^{\perp}}= \sum_q b_q^t\tilde \bmu_q$ are all of the same sign and satisfy \eqref{eq:n10}. Without loss of generality we may assume that all $b_q^t$ are non-negative.
The numbers
\[
l_t= l(b_1^t, \dots, b_{K+ 1}^t):= 2L_{m+ 1}\sum_{q}b_q^t+ 2\lu \mathbf X, (\bth_v)_{\GV}\ru+ 2\lu \bphi_v, \bth_v\ru+ |\bth_v|^2
\]
satisfy
$\rho_n^{\al_{m+1}}\rho_n^{0-}\lesssim l_t\lesssim  \rho_n^{\al_{m+1}}\rho_n^{0+}$, since
\[
\big|2\lu \mathbf X,(\bth_v)_{\GV}\ru+ 2\lu \bphi_v, \bth_v\ru+ |\bth_v|^2\big|\lesssim \rho_n^{\al_m+ 0+}.
\]
This numbers depend on $\mathbf X$, but not on $\mathbf\Phi$ or $\rho$. The numbers $k_t= k(b_1^t, \dots, b_{K+ 1}^t)$ are positive, integer, and independent of $\bxi$.

The following lemma is identical to Lemma 10.4 of \cite{ParSht2}, where for our purposes we have replaced the explicit constants $1/2$ and $2/3$ by $\vartheta$ and $\varsigma$, respectively.

\bel\label{lem:integral1}
For $1\leqslant K\leqslant d-1$; $n_1, \dots, n_{K+ 1}\in \N_0$; $k_1, \dots, k_T\in \N$ let $Q:= \sum_{t= 1}^Tk_t$,
\begin{equation*}%\label{eq:in2}
\hat J_K:= \int\limits_{\CM_p}\frac{(\sin\Phi_1)^{n_1}\dots (\sin\Phi_K)^{n_K}(\sin\Phi_{K+ 1})^{n_{K+ 1}}\,d\mathbf\Phi}{\prod_{t= 1}^T \big(l_t+ \rho\sum_{j= 1}^{K+ 1} b_j^t \sin\Phi_j\big)^{k_t}
}.
\end{equation*}
Then there exist positive numbers $\delta_0$, $p_K$, and $q_K$ depending only on the constants \eqref{beta and alphas} and $K$ such that
\begin{equation*}%\label{eq:JK3}
\hat J_K= \sum_{q= 0}^K(\ln\rho)^q\sum_{p= 0}^\infty
e(p,q){\rho}^{-p},
\end{equation*}
where
\bees%\label{eq:estimatee1}
\big|e(p,q)\big|\lesssim\rho_n^{(\varsigma- p_K)p}\rho_n^{-Q\beta}.
\enes
These estimates are uniform in the following regions of variables:
\bees%\label{eq:range1}
\rho_n^{\beta}\lesssim l_t\lesssim\rho_n^{\vartheta},\ \ \rho_n^{-\delta_0}\lesssim b_j^t\lesssim \rho_n^{\delta_0},\ \ \rho_n^{\varsigma- q_K}<{\rho}.
\enes
\enl

Now using Lemma~\ref{lem:integral1} we can compute the integrals of \eqref{S^f} over the domain $\{\mathbf\Phi\in \CM_p\}$ (recall that this integration is not needed for $K= 0$ by Remark~\ref{K= 0 remark}). Substituting the result into \eqref{integration in Phi}, integrating in $d\bnu$ over $\Omega(\GV)$, and taking into account \eqref{eq:n4} and \eqref{eq:n6} we obtain in the region $2\rho_n/3< \rho< 6\rho_n$
\begin{equation*}%\label{direct answer}
\begin{split}
&\vol\CA^+_p(\rho)- \vol\CA^-_p(\rho)\\ &= \sum_{q= 0}^K\sum_{h= 0}^{(l_0+ 1)\tilde k}\sum_{\iota_1, \dots, \iota_h\in \widetilde J}\sum_{j= 0}^\infty C_{q\, h\, j}^{\iota_1\cdots \iota_h}\rho^{K+ 1+ (2- 2w)h+ \iota_1+ \cdots+ \iota_h- j}(\ln\rho)^q+ O(\rho_n^{-M+ 2w- d}),
\end{split}
\end{equation*}
with the coefficients satisfying
\begin{equation*}%\label{direct coefficients}
|C_{q\, h\, j}^{\iota_1\cdots \iota_h}|\lesssim \rho_n^{-2\b h+ \varsigma j}.
\end{equation*}
This, together with equations \eqref{eq:n2}, \eqref{eq:densityh3}, Lemma~\ref{H1H2 lemma}, relation \eqref{beta and alphas}, Section~11 of \cite{ParSht2}, and the observation that the number of different quasi-lattice subspaces $\GV$ is $\lesssim\rho_n^{0+}$, completes the proof of Lemma \ref{main_lem} and, thus, of our main theorem in the case of $B= \widetilde B$ with the symbol satisfying \eqref{tilde b}. As explained at the end of Section~\ref{reduction section}, the summation over $\widetilde J$ may be replaced by summation over $J_0$.

It remains to relax the assumptions on $B$. This will be done in the subsequent section.

\section{Approximation}\label{final section}

In this section we prove Lemma~\ref{main_lem} and thus Theorem~\ref{main_thm} for general $B$ using the fact that the proof is complete for $\widetilde B$ whose symbol fulfills the extra assumption \eqref{tilde b}.

\subsection*{1.}
Given $B$ satisfying the hypothesis of Theorem~\ref{main_thm} and the number $M$, we fix the values of $k$ and $\tilde k$ in such a way that Lemma~\ref{main_lem} holds true for $H= (-\Delta)^w+ \widetilde B$, where the symbol $\tilde b$ of $\widetilde B$ satisfying \eqref{tilde b} is constructed at the end of Section~\ref{reduction section}.
For $R> 0$ let us define (recall \eqref{CP})
\begin{equation*}
\CP_R:= \CP^L(\CB_{R}), \quad \CP_R^c:= \CP^L(\Rd\setminus \CB_{R})
\end{equation*}

We start by estimating the quadratic form of $B- \widetilde B$. For any $\psi\in \plainH{2w}(\R^d)$
\begin{equation}\label{B correction estimate}
\begin{split}
\big|\langle\psi, (B- \widetilde B)\psi\rangle\big|&\leqslant \big|\langle\psi, \CP_{R_0}(B- \widetilde B)\CP_{R_0}\psi\rangle\big|+ \big|\langle\psi, \CP_{R_0}(B- \widetilde B)\CP_{R_0}^c\psi\rangle\big|\\ &+ \big|\langle\psi, \CP_{R_0}^c(B- \widetilde B)\CP_{R_0}\psi\rangle\big|+ \big|\langle\psi, \CP_{R_0}^c(B- \widetilde B)\CP_{R_0}^c\psi\rangle\big|.
\end{split}
\end{equation}
By Condition \eqref{eq:condB2}, the symbol of $(B- \widetilde B)\CP_{R_0}^c$ satisfies
\begin{equation*}
{\,\vrule depth4pt height11pt width1pt}\,(b- \tilde b)\Id_{\Rd\setminus \CB_{R_0}}{\vrule depth4pt height11pt width1pt\,}_{\vark/2,\, 0}^{(\vark/\beta)}< \rho_n^{-k}.
\end{equation*}
Now Propositions~\ref{bound:prop} and \ref{product:prop} imply that
\begin{equation}\label{bound from Condition B}
\big\|(-\Delta+ 1)^{-\vark/4}(B- \widetilde B)(-\Delta +1)^{-\vark/4}\CP_{R_0}^c\big\|\leqslant C\rho_n^{-k}.
\end{equation}
Hence
\begin{equation}\label{one term estimated}
\begin{split}
&\big|\langle\psi, \CP_{R_0}(B- \widetilde B)\CP_{R_0}^c\psi\rangle\big|\\ &= \big|\langle(-\Delta+ 1)^{\vark/4}\psi, \CP_{R_0}(-\Delta+ 1)^{-\vark/4}(B- \widetilde B)\CP_{R_0}^c(-\Delta+ 1)^{-\vark/4}(-\Delta+ 1)^{\vark/4}\psi\rangle\big|\\ &\leqslant C\rho_n^{-k}\langle\psi, (-\Delta+ 1)^{\vark/2}\psi\rangle,
\end{split}
\end{equation}
and the analogous estimates hold for the last two terms in \eqref{B correction estimate}. Thus \eqref{B correction estimate} implies
\begin{equation}\label{estimate with B^(k)}
|B- \widetilde B|\leqslant B^{(k)},
\end{equation}
where $B^{(k)}$ is the operator of multiplication by the function
\begin{equation}\label{b^(k)}
b^{(k)}(\bxi):= \begin{cases}\|b\|_{\plainL{\infty}(\Rd\times\CB_{R_0})}+ \|\tilde b\|_{\plainL{\infty}(\Rd\times\CB_{R_0})}, &|\bxi|\leqslant R_0,\\ C\rho_n^{-k}\big(1+ |\bxi|^2\big)^{\vark/2}, &|\bxi|> R_0\end{cases}
\end{equation}
in the momentum space.

In view of Lemma~\ref{norms lemma}(a), we conclude that
\begin{equation}\label{up and down}
N\big((-\Delta)^w+ B, \lambda\big)\gtrless N\big((-\Delta)^w+ \widetilde B \pm B^{(k)}, \lambda\big).
\end{equation}
So to prove \eqref{eq:main_lem1} it will be sufficient to show that for $\rho\in I_n$ (which we assume everywhere below) the right hand side of \eqref{up and down} does not differ from $N\big((-\Delta)^w+ \widetilde B, \rho^{2w}\big)$ by more than $O(\rho_n^{-M})$.
By \eqref{eq:main_lem1} and Remark~\ref{spurious remark}, it is enough to prove that
\begin{equation}\label{approximation goal}
N\big((-\Delta)^w +\widetilde B \pm B^{(k)}, \lambda\big) =N\big((-\Delta)^w +\widetilde B, \lambda +O(\rho_n^{2w -d -M})\big).
\end{equation}

\subsection*{2.}
We note that for
\begin{equation}\label{R_*}
R_*:= (4\rho_n^{d+ M})^{1/(w- \vark)}
\end{equation}
we have
\begin{equation}\label{intermediate irrelevant}
N\big((-\Delta)^w+ \widetilde B \pm B^{(k)}, \lambda\big)= N\big((-\Delta)^w+ \widetilde B \pm\CP_{R_0}B^{(k)} \pm\CP_{R_*}^cB^{(k)}, \lambda+ O(\rho_n^{2w -d -M})\big).
\end{equation}
Indeed,
\begin{equation*}
\big\|(\CP_{R_*}- \CP_{R_0})B^{(k)}\big\|= C\rho_n^{-k}(1+ R_*^2)^{\vark/2}= O(\rho_n^{2w -d -M})
\end{equation*}
in view of \eqref{k fist condition}.

\subsection*{3.}
Now we are going to prove that
\begin{equation}\label{interior removal}
 N\big((-\Delta)^w+ \widetilde B \pm\CP_{R_0}B^{(k)} \pm\CP_{R_*}^cB^{(k)}, \lambda\big)= N\big((-\Delta)^w+ \widetilde B \pm\CP_{R_*}^cB^{(k)}, \lambda+ O(\rho_n^{2w -d -M})\big).
\end{equation}
This will be done with the help of the following lemma, which is a development of Lemma~3.1 from \cite{Par}.

\bel\label{iterative decay lemma}
Let $H_0$, $V$, $A$ be pseudo--differential operators with almost--periodic coefficients. Suppose that $H:= H_0+ V$ is elliptic, selfadjoint and bounded below, and there exists a collection of orthogonal projections $\{P_l\}_{l= 0}^L$ commuting with $H_0$ such that
\begin{equation}\label{projector conditions}
\sum_{l =0}^LP_l =I \quad \textrm{and} \quad V_{n\, l} :=P_nVP_l =0 \quad \textrm{for} \quad |l -n|> 1.
\end{equation}
Suppose that $A= P_0A$ and that
\begin{equation*}
a:= \|A\|< \infty.
\end{equation*}
At last, suppose that for $\lambda \in\R$
\begin{equation}\label{D_l hypothesis}
D_l:= \dist\big(\lambda, \sigma(P_lHP_l)\big) -(4+ 2^{5 -L})a >0, \quad l =0, \dots, L -1
\end{equation}
and
\begin{equation}\label{V assumption}
\max_{0 \leqslant l \leqslant L -1}\big(a+ \|V_{l\, l -1}\|+ \|V_{l\, l+ 1}\|\big)/D_l\leqslant 1/4.
\end{equation}
Then for
\begin{equation}\label{vareps}
\varepsilon:= 2^{4 -L}a
\end{equation}
we have
\begin{equation}\label{N(H+ A)}
N(H, \lambda -\varepsilon) \leqslant N(H+ A, \lambda) \leqslant N(H, \lambda +\varepsilon).
\end{equation}
\enl

\bep
We will prove the first inequality; the second follows by interchanging the roles of $H_0$ and $H_0 +A$. Let $E_\lambda$ be the spectral projection of $(-\infty, \lambda]$ for $H$. By Lemma~4.1 of \cite{ParSht2} it is enough to prove that
\begin{equation}\label{lambda form bound}
\langle \phi, (H+ A)\phi\rangle\leqslant \lambda\|\phi\|^2 \quad \textrm{for every}\quad \phi \in E_{\lambda- \varepsilon}\plainL2(\Rd).
\end{equation}
Let
\begin{equation}\label{delta and K}
\delta :=\min\{a, 2^{-3 -L}\min_{0 \leqslant l \leqslant L -1}D_l\}, \quad K :=[2a/\delta] +2,
\end{equation}
so that
\begin{equation}\label{K -1 bounds}
2a \leqslant (K -1)\delta \leqslant 3a
\end{equation}
and by \eqref{V assumption}
\begin{equation}\label{K -1 estimate}
K -1\leqslant 3a/\delta \leqslant3\max\{1, 2^{L +3}a/\min_{0 \leqslant l \leqslant L -1}D_l\} \leqslant 2^{L +3}.
\end{equation}

For $\phi \in E_{\lambda- \varepsilon}$ introduce
\begin{equation*}
\begin{split}
\phi^k &:=(E_{\lambda -\varepsilon -(k -1)\delta} -E_{\lambda -\varepsilon -k\delta})\phi, \quad k =1, \dots, K -1, \\ \phi^K&:= E_{\lambda -\varepsilon -(K- 1)\delta}\phi, \quad \phi' :=\phi -\phi^K =\sum_{k =1}^{K -1}\phi^k.
\end{split}
\end{equation*}
Then $\phi =\sum_{k =1}^K\oplus\phi^k$ and, letting
\begin{equation}\label{eta^k}
\eta^k :=H\phi^k -\big(\lambda -\varepsilon -(k -1)\delta\big)\phi^k, \quad k =1, \dots, K -1,
\end{equation}
we have
\begin{equation}\label{eta norm}
\|\eta^k\| \leqslant\delta\|\phi^k\|.
\end{equation}
Let $P_{-1} := P_{L +1} := 0$. Projecting \eqref{eta^k} with $P_l$ we obtain
\begin{equation*}
\eta_l^k= V_{l\, l -1}\phi_{l -1}^k+ \Big(P_lHP_l- \big(\lambda -\varepsilon -(k -1)\delta\big)\Big)\phi^k_l+ V_{l\, l +1}\phi^k_{l +1}, \quad l =0, \dots, L,
\end{equation*}
and thus by \eqref{V assumption}, \eqref{delta and K} and \eqref{eta norm}
\begin{equation*}
\begin{split}
\|\phi^k_l\| &\leqslant \big(\|\eta^k_l\| +\|V_{l\, l -1}\|\|\phi_{l -1}^k\| +\|V_{l\, l +1}\|\|\phi_{l +1}^k\|\big)/D_l \\ &\leqslant 2^{-3 -L}\|\phi^k\| +\|\phi^k_{l -1}\|/4 +\|\phi^k_{l +1}\|/4, \quad l =0, \dots, L -1.
\end{split}
\end{equation*}
By induction, starting from $l =0$ we obtain
\begin{equation*}
\|\phi^k_l\| \leqslant 2^{-2 -L}\|\phi^k\| +3\|\phi^k_{l +1}\|/8, \quad l =0, \dots, L -1.
\end{equation*}
Again by induction, using that $\|\phi^k_L\| \leqslant\|\phi^k\|$, we get $\|\phi^k_l\| \leqslant2^{l -L}\|\phi^k\|$, $l =1, \dots, L$ and thus $\|\phi^k_0\| \leqslant2^{-L}\|\phi^k\|$. Therefore, for $k =1, \dots, K -1$,
\begin{equation*}
\|A\phi^k\| =\|A\phi^k_0\| \leqslant2^{-L}a\|\phi^k\|,
\end{equation*}
and thus
\begin{equation*}
\|A\phi'\|\leqslant \sum_{k =1}^{K -1}\|A\phi^k\|\leqslant 2^{-L}\sqrt{K -1}a\|\phi'\|
\end{equation*}
and
\begin{equation*}
\big|\lu\phi', A\phi'\ru\big| =\Big|\sum_{k, m =1}^{K -1}\lu\phi^k_0, A\phi^m_0\ru\Big| \leqslant 2^{-2L}(K -1)a\|\phi'\|^2.
\end{equation*}
Hence
\begin{equation*}
\begin{split}
\lu\phi, (H +A)\phi\ru &=\lu\phi', H\phi'\ru +\lu\phi', A\phi'\ru +2\Re\lu\phi^K, A\phi'\ru +\lu\phi^K, H\phi^K\ru +\lu\phi^K, A\phi^K\ru\\ &\leqslant (\lambda- \varepsilon)\|\phi'\|^2 +2^{-2L}(K -1)a\|\phi'\|^2+ 2^{1 -L}\sqrt{K -1}a\|\phi'\|\|\phi^K\|\\ &+ \big(\lambda -\varepsilon -(K -1)\delta\big)\|\phi^K\|^2+ a\|\phi^K\|^2\\ &\leqslant\big(\lambda -\varepsilon +2^{1 -2L}(K -1)a\big)\|\phi'\|^2 +\big(\lambda -\varepsilon -(K -1)\delta +2a\big)\|\phi^K\|^2\\ &\leqslant\lambda\|\phi\|^2,
\end{split}
\end{equation*}
where the last inequality follows from \eqref{K -1 bounds} and \eqref{K -1 estimate}.
\enp

We now want to apply Lemma~\ref{iterative decay lemma} to
\begin{equation*}
H_0^\pm :=(-\Delta)^w \pm\CP_{R_*}^cB^{(k)}, \quad V :=\widetilde B, \quad A^\pm :=\pm\CP_{R_0}B^{(k)}.
\end{equation*}
Note that
\begin{equation}\label{a}
a :=\|b\|_{\plainL{\infty}(\Rd\times\CB_{R_0})}+ \|\tilde b\|_{\plainL{\infty}(\Rd\times\CB_{R_0})}
\end{equation}
does not depend on $\rho_n$. For
\begin{equation}\label{L}
L :=\big[4 +\log_2a +(M +d -2w)\log_2\rho_n\big] +1
\end{equation}
we let
\begin{equation}\label{R_l}
R_l :=R_0 +l\rho_n^{2/k}, \quad l =0, \dots, L -1,
\end{equation}
and introduce a family of projections
\begin{equation}\label{P_l}
P_0 :=\CP_{R_0}, \quad P_l :=\CP_{R_l} -\CP_{R_{l -1}}, \quad l =1, \dots, L -1, \quad P_L :=\CP_{R_{L -1}}^c.
\end{equation}
Let us check that the hypothesis of Lemma~\ref{iterative decay lemma} is satisfied. Relation \eqref{projector conditions} follows from \eqref{R(rho)} and \eqref{R_l}. It follows from \eqref{tilde B estimate} that for $l\leqslant L -1$
\begin{equation}\label{norm on P_l}
\|P_lHP_l\| \leqslant\Big\|P_{L -1}\big((-\Delta)^w+ \widetilde B\big)P_{L -1}\Big\| \leqslant 2\big\|P_{L -1}(-\Delta)^wP_{L -1}\big\| \leqslant 2(R_{L -1})^{2w}.
\end{equation}
Also, for $l\leqslant L -1$
\begin{equation}\label{norms of projected V}
\|V_{l\, l -1}\|+ \|V_{l\, l +1}\| \leqslant 2(R_{L -1} +\rho_n^{2/k})^{2\tilde w}.
\end{equation}
Since by \eqref{L} and \eqref{R_l} we have
\begin{equation}
R_{L -1} =R_0 +\big[4 +\log_2a +(M +d -2w)\log_2\rho_n\big]\rho_n^{2/k}\lesssim \rho_n^{2/k}\log\rho_n,
\end{equation}
relations \eqref{D_l hypothesis} and \eqref{V assumption} follow from \eqref{norm on P_l} and \eqref{norms of projected V} if $\rho_n$ is big enough.

Applying Lemma~\ref{iterative decay lemma}, we get \eqref{N(H+ A)} with
\begin{equation*}
\varepsilon =2^{4- L}a \leqslant\rho_n^{-M +2w -d},
\end{equation*}
which implies \eqref{interior removal}.

\subsection*{4.}
It remains to prove that
\begin{equation}\label{exterior removal}
N\big((-\Delta)^w+ \widetilde B \pm\CP_{R_*}^cB^{(k)}, \lambda\big)= N\big((-\Delta)^w+ \widetilde B, \lambda+ O(\rho_n^{2w -d -M})\big).
\end{equation}
Choose
\begin{equation}\label{vareps again}
\varepsilon :=\rho_n^{-d -M}.
\end{equation}
In view of \eqref{tilde B estimate}, we have
\begin{equation*}
(-\Delta)^{\tilde w} +\widetilde B \lessgtr\CP_{R_*}(1 \pm\varepsilon)\big((-\Delta)^{\tilde w} +\widetilde B\big)\CP_{R_*} \oplus\CP_{R_*}^c(1 \pm1/\varepsilon)\big((-\Delta)^{\tilde w} +\widetilde B\big)\CP_{R_*}^c.
\end{equation*}
Therefore,
\begin{equation}\label{R_* direct sum}
\begin{split}
(-\Delta)^w+ \widetilde B \pm\CP_{R_*}^cB^{(k)} &\lessgtr\CP_{R_*}\big((-\Delta)^w \pm\varepsilon(-\Delta)^{\tilde w} +(1 \pm\varepsilon)\widetilde B\big)\CP_{R_*} \\ &\oplus\CP_{R_*}^c\big((-\Delta)^w \pm(-\Delta)^{\tilde w}/\varepsilon +(1 \pm1/\varepsilon)\widetilde B \pm B^{(k)}\big)\CP_{R_*}^c.
\end{split}
\end{equation}
Using \eqref{tilde B estimate} again and recalling the definitions \eqref{vareps again}, \eqref{b^(k)} and \eqref{R_*}, we can estimate the last term on the right hand side of \eqref{R_* direct sum} from below:
\begin{equation*}
\begin{split}
&\CP_{R_*}^c\big((-\Delta)^w \pm(-\Delta)^{\tilde w}/\varepsilon +(1 \pm1/\varepsilon)\widetilde B \pm B^{(k)}\big)\CP_{R_*}^c \\ &>\big((-\Delta)^w -2(-\Delta)^{\tilde w}/\varepsilon\big)\CP_{R_*}^c \geqslant (R_*^{2w} -2R_*^{2\tilde w}/\varepsilon)\CP_{R_*}^c \geqslant (5\rho_n)^{2w}\CP_{R_*}^c,
\end{split}
\end{equation*}
so it does not contribute to the density of states for $\rho \in I_n$.
For the first term we have
\begin{equation*}
\CP_{R_*}\big((-\Delta)^w \pm\varepsilon(-\Delta)^{\tilde w} +(1 \pm\varepsilon)\widetilde B\big)\CP_{R_*} \lessgtr\CP_{R_*}(1 \pm\varepsilon)\big((-\Delta)^w +\widetilde B\big)\CP_{R_*},
\end{equation*}
so
\begin{equation}\label{bound with lambda rescaled}
\begin{split}
&N\Big((-\Delta)^w+ \widetilde B \pm\CP_{R_*}^cB^{(k)}, \lambda\Big)\\
&\gtrless N\Big(\CP_{R_*}\big((-\Delta)^w \pm\varepsilon(-\Delta)^{\tilde w} +(1 \pm\varepsilon)\widetilde B\big)\big|_{\CP_{R_*}\plainL{2}(\Rd)}, \lambda\Big)\\ &\gtrless N\Big(\CP_{R_*}\big((-\Delta)^w +\widetilde B\big)\big|_{\CP_{R_*}\plainL{2}(\Rd)}, \lambda/(1 \pm\varepsilon)\Big),
\end{split}
\end{equation}
and the same estimates hold true for $B^{(k)}$ replaced by $0$. Combining these two versions of \eqref{bound with lambda rescaled}, we obtain
\begin{equation}\label{difference of IDS}
\begin{split}
&N\big((-\Delta)^w+ \widetilde B, \lambda\big)\lessgtr N\big((-\Delta)^w+ \widetilde B \mp\CP_{R_*}^cB^{(k)}, \lambda\big)\\ &\lessgtr N\Big(\CP_{R_*}\big((-\Delta)^w +\widetilde B\big)\big|_{\CP_{R_*}\plainL{2}(\Rd)}, \lambda/(1 \mp\varepsilon)\Big) \lessgtr N\big((-\Delta)^w+ \widetilde B, (1 \pm\varepsilon)\lambda/(1 \mp\varepsilon)\big).
\end{split}
\end{equation}
Recalling that $\lambda =\rho^{2w} \leqslant(4\rho_n)^{2w}$ and \eqref{vareps again}, we arrive at \eqref{exterior removal}.

Combining \eqref{intermediate irrelevant}, \eqref{interior removal} and \eqref{exterior removal}, we get \eqref{approximation goal}.

\end{document}